\documentclass[review]{elsarticle}

\usepackage{amssymb}
\usepackage{apalike}
\usepackage{lineno,hyperref}
\usepackage{graphicx}
\usepackage{xcolor,stfloats}
\usepackage{subfigure}
\usepackage{amsmath}
\usepackage{algorithm}
\usepackage{algorithmic}
\usepackage{booktabs} 
\usepackage{graphicx}
\usepackage{xcolor,stfloats}
\usepackage{stfloats}
\usepackage{verbatim}
\usepackage{amscd}
\usepackage{url}
\usepackage{enumitem} 
\usepackage{amsfonts}
\usepackage{dcolumn}
\usepackage{tabularx} 
\usepackage{setspace}
\usepackage{bm}
\usepackage{multirow}
\usepackage{textcomp}
\usepackage{threeparttable}
\usepackage{setspace}
\usepackage[figuresright]{rotating}
\usepackage{color}
\usepackage{arydshln}
\usepackage{hhline}
\usepackage{times}

\journal{Expert Systems with Applications}
\biboptions{authoryear}

\begin{document}
	\begin{frontmatter}
		
		
		\begin{titlepage}
			\begin{center}
				\vspace*{1cm}
				
				\textbf{ \large Towards a performance characteristic curve for model evaluation: an application in information diffusion prediction}
				
				\vspace{1.5cm}
				
				Wenjin Xie$^{a}$ (xiewenjin@email.swu.edu.cn), Xiaomeng Wang$^a$ (wxm1706@swu.edu.cn), Rados{\l}aw Michalski$^b$ (radoslaw.michalski@pwr.edu.pl), Tao Jia$^a$ (tjia@swu.edu.cn) \\
				
				\hspace{10pt}
				
				\begin{flushleft}
					\small  
					$^a$ College of Computer and Information Science, Southwest University, Chongqing, 400037, China \\
					$^b$ Department of Artificial Intelligence, Faculty of Information and Communication Technology, Wrocław University of Science and Technology, Wrocław, 50-370, Poland

					\vspace{1cm}
					\textbf{Corresponding author:} \\
					Xiaomeng Wang, Tao Jia \\
					College of Computer and Information Science, Southwest University, Chongqing, 400037, China \\
					Email: wxm1706@swu.edu.cn, tjia@swu.edu.cn
					
				\end{flushleft}        
			\end{center}
		\end{titlepage}
		
		\title{Towards a performance characteristic curve for model evaluation: an application in information diffusion prediction}
		
		\author[label1]{Wenjin Xie}
		\ead{xiewenjin@email.swu.edu.cn}
		
		\author[label1]{Xiaomeng Wang \corref{cor1}}
		\ead{wxm1706@swu.edu.cn}

        \author[label2]{Rados{\l}aw Michalski}
		\ead{radoslaw.michalski@pwr.edu.pl}
		
		\author[label1]{Tao Jia \corref{cor1}}
		\ead{tjia@swu.edu.cn}
		
		\cortext[cor1]{Corresponding author.}
		\address[label1]{College of Computer and Information Science, Southwest University, Chongqing, 400037, China}
		\address[label2]{Department of Artificial Intelligence, Faculty of Information and Communication Technology, Wrocław University of Science and Technology, Wrocław, 50-370, Poland}
		
		\begin{abstract}
			The information diffusion prediction on social networks aims to predict future recipients of a message, with practical applications in marketing and social media. While different prediction models all claim to perform well, general frameworks for performance evaluation remain limited. Here, we aim to identify a performance characteristic curve for a model, which captures its performance on tasks of different complexity. We propose a metric based on information entropy to quantify the randomness in diffusion data. We then identify a scaling pattern between the randomness and the prediction accuracy of the model. By properly adjusting the variables, data points by different sequence lengths, system sizes, and randomness can all collapse into a single curve. The curve captures a model's inherent capability of making correct predictions against increased uncertainty, which we regard as the performance characteristic curve of the model. The validity of the curve is tested by three prediction models in the same family, reaching conclusions in line with existing studies. In addition, we apply the curve to successfully assess the performance of eight state-of-the-art models, providing a clear and comprehensive evaluation even for models that are challenging to differentiate with conventional metrics. Our work reveals a pattern underlying the data randomness and prediction accuracy. The performance characteristic curve provides a new way to evaluate models' performance systematically, and sheds light on future studies on other frameworks for model evaluation.
		\end{abstract}
		
		
	\end{frontmatter}

\section{Introduction}

In machine learning, a model can be regarded as a collection of the algorithm, the parameters and other things that can recognize a certain pattern in the data and utilize the pattern to forecast something unknown \citep{Schelter2015}. For a given task, there are usually multiple models, which naturally gives rise to a question on how to effectively evaluate them \citep{raschka2018model}. Indeed, all models are claimed to outperform the existing baselines when proposed. However, systematic comparisons are limited and a general framework for the performance evaluation remains to be explored. It is usually unclear if a model's reported advances only hold in a parameter region delicately selected or one model absolutely outperforms other baseline models in all cases.

In traditional engineering fields, the properties of instruments are usually evaluated in a more comprehensive way. For example, the performance of an engine can be evaluated by the performance characteristic curve which illustrates how the output power and torque vary with the engine's rotation speed. In this way, the parameter region that is optimal for an engine can be identified. Two engines can be properly compared and selected for different tasks. This motivates us to seek the performance characteristic curve for a machine learning model, which tells how the performance of the model changes with different complexity of the task. Nevertheless, plotting such a performance curve is not trivial. The selection of physical quantity to gauge the complexity is not straightforward. As will be shown later, when simply plotting the model's output with some kind of complexity measure of the data, the data points will be scattered and a consistent relationship is hardly guaranteed.

In this paper, we take the evaluation task of information diffusion prediction on social networks as a particular case. Information diffusion on social networks has drawn massive attention due to the rapid development of social media. These social media (such as X, Weibo, TikTok, etc.) have become one of the major pathways to share and exchange information and ideas, which profoundly shapes the contemporary social, economic and political environment \citep{auxier2021social, zhang2022investigation}. Consequently, there is an urgent need to understand patterns underlying information diffusion in social media, which helps forecast a message's future impact and control the potential hazard that comes after. Intensive research has been carried out from different aspects by researchers in diverse fields such as computer science, physics, mathematics and social science. To simulate the spreading process, different theoretical models are proposed, from the susceptible-infectious-recovered (SIR) model in epidemics \citep{hethcote1989three}, to independent cascade (IC) model frequently used in computer science \citep{kempe2003maximizing}, and to linear threshold (LT) model \citep{watts2002simple} that takes social reinforcement effects into consideration. In the interdisciplinary field of network science, researchers are interested in identifying central nodes or topological features that could maximize the influence \citep{kempe2003maximizing}, building connections between information spreading and statistical physics process in order to control or promote the spreading \citep{xie2021detecting, michalski2022temporal}. 

\begin{figure}[t]
	\centering
	\includegraphics[width=0.9\textwidth]{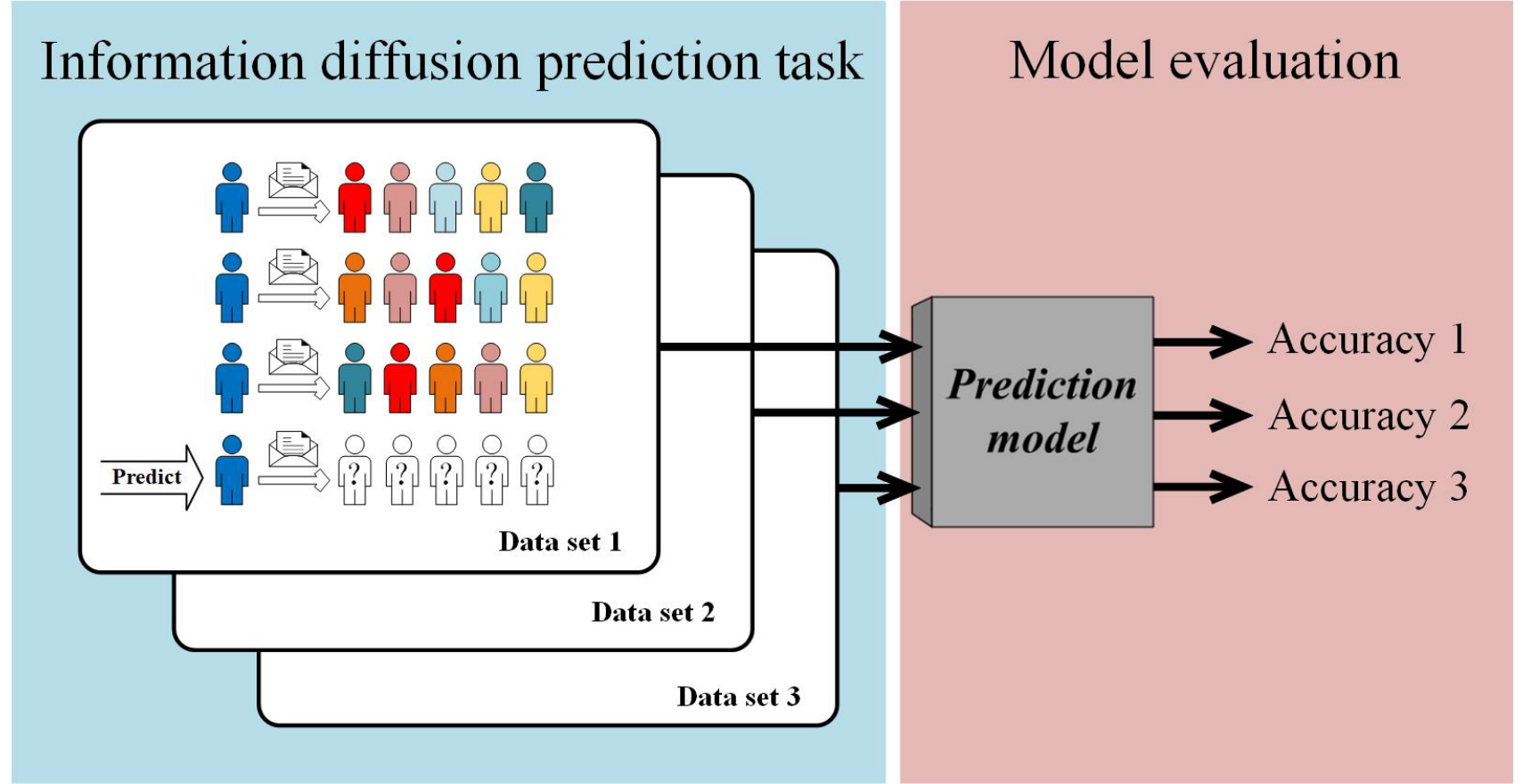}
	\caption{A simple example illustrating the information diffusion prediction task and the current commonly used method of evaluating the prediction model, which is simply calculating the model's accuracy on the single-point metric for each discrete data set. \label{fig:intro}}
\end{figure}\noindent

Here, we specifically focus on predicting the set of nodes (users) that are reached by the information, given the nodes that initiate the spreading. Compared to the macro-level information diffusion prediction tasks that focus on predicting the scope and scale of information spread, this task pays more attention to the specific individuals affected by the information and their affected time. This kind of task is therefore considered as the micro-level information diffusion prediction task correspondingly \citep{yang2019neural, wang2023multiscale}.
A lot of external features can be applied in this prediction task, such as the network topology serving as the backbone of the spreading, the information content, and the precise time of message acceptance \citep{wang2017topological, wang2023multiscale, liu2023information}. To avoid the interference caused by multiple features, we consider the simplest approach that makes use of the current ordered sequence of recipients to predict future recipients. This kind of models learns from the full spreading sequence in the training set. Based on patterns learned, the model predicts the successive nodes once an initiating node is given. In the following text, these models will serve as application objects for our model evaluation framework. With the tools in deep learning and network embedding, these models significantly advance the front line of this research field. But as mentioned, while all proposed models claim to be better than other baselines, systematic comparisons remain limited. The common benchmark data set should be a beneficial solution for the model comparison, as researchers have done in fields like computer vision. However, the task of information diffusion prediction is widely recognized as lacking common data sets in general \citep{hofman2017prediction, martin2016exploring}. Researchers usually collect data on their own and simply evaluate prediction models on these isolated data sets by their prediction accuracy, as shown in Figure \ref{fig:intro}. Hence, even for data collected from the same social platform, they could differ due to different collection time, topics, the set of users covered, relationships extracted, and collection methods applied. Although there are certain public data sets widely used in the field, the different filtering criteria and data pre-processing would still make the information eventually applied in one study different from the other. All these uncontrollable variables bring ambiguities to performance comparison in information diffusion. In this paper, we show that the performance characteristic curve of a model can be reached by leveraging the scaling pattern between the inherent randomness in the data and the accuracy of the prediction. Furthermore, we propose a new framework for model evaluation using the performance characteristic curve.

The major contribution of this paper can be summarized as follows:
\begin{enumerate}
\item We define a metric named \underline{A}verage \underline{P}airwise \underline{C}omparison \underline{E}ntropy (APCE) for quantifying the randomness of the sequential data based on the information entropy. APCE captures the order information of every pair of nodes in the information diffusion sequence data.

\item We discover the scaling pattern between the data randomness and the performance of prediction models which fits well with an exponentially decreasing curve. We take it as the performance characteristic curve of the model, and employ the curve to evaluate the predictive performance of the machine learning model.

\item We conduct experiments on 3540 synthetic and empirical data sets to validate the effectiveness of our evaluation method. Furthermore, we prove the advantages of our method through an accurate and comprehensive evaluation of eight state-of-the-art models, which cannot be clearly distinguished with the existing metrics.

\end{enumerate}

The rest of this paper is organized as follows. In Section \ref{sec:related_studies} related studies are reviewed on the model evaluation, the data randomness, and the relationship between randomness and the model performance. Section \ref{sec:pre} introduces the definition of information diffusion prediction task. The commonly used metric for the model evaluation is also introduced and its limitations are analyzed. Section \ref{sec:method} proposes our solution for the model evaluation which is based on the designed metric for data randomness and the scaling pattern between data randomness and model performances we identify. In Section \ref{sec:experiments} we conduct experiments on both synthetic and empirical data. The experimental results validate the effectiveness the performance characteristic curve and further demonstrate its prominence and comprehensiveness over the existing evaluation metrics. Section \ref{sec:conclusion} concludes the paper and proposes some suggestions for future works.

\section{Related studies} \label{sec:related_studies}
In this section, we review the relevant literature of two key aspects of our work: studies on metrics for model evaluation and studies on quantifying the data randomness and associating it with model performance.

\subsection{Metrics for model evaluation}
In machine learning, numerous metrics are proposed to evaluate model performance across different tasks.  For classification and regression, metrics include Precision, Recall, Mean Squared Error (MSE), and Mean Absolute Error (MAE); for retrieval and recommendation tasks, Discounted Cumulative Gain (DCG), Normalized Cumulative Gain (NCG), and Diversity are standard \citep{su1992evaluation, kunaver2017diversity, raschka2018model, rainio2024evaluation}.

These traditional evaluation metrics are often applied to assess models across various isolated data sets. However, the difficulty of tasks across different data sets can vary significantly, due to factors such as class imbalance, noise in the samples, and differences in data scale. Consequently, evaluating model capabilities solely based on performance across these isolated data sets can be misleading. To better evaluate a model's generalization ability and performance across tasks of varying complexity, researchers have commenced the incorporation of task difficulty into the design and evaluation of models. ~\cite{bengio2009curriculum} quantify the difficulty of a task by the amount of noise in data or the variability and complexity of geometric shapes in a graph task, then let the model learn from simple tasks and gradually introduce more difficult tasks.  ~\cite{zamir2018taskonomy} investigate the relationships between different tasks and proposes knowledge transfer based on task difficulty. The research shows that tasks can be hierarchically organized according to their complexity and difficulty, allowing for the optimization of a model’s learning path in multi-task environments. ~\cite{pentina2014pac} introduce a method for evaluating model generalization in the context of lifelong learning by considering the difficulty and complexity of different tasks.

These studies indicate that the machine learning community is gradually recognizing the limitations of model performance on isolated data sets and is exploring more sophisticated methods to design and assess models. These efforts aim to provide a more accurate reflection of the comprehensive performance of models.

\subsection{Quantifying the randomness of data}
The extent of randomness in the data is usually quantified by information entropy \citep{shannon1948mathematical}. ~\cite{mackay2003information} emphasizes the important role of information entropy in measuring uncertainty in Bayesian inference. ~\cite{liu2014no} utilize the spatial and spectral information entropy to assess the quality of image data. In the field of natural language processing, information entropy is used to analyze the vocabulary and grammatical structure in text data, in order to measure the complexity and uncertainty of language \citep{berger1996maximum}. In the field of network security, information entropy also plays an important role in identifying potential network attacks due to its capability to detect abnormal patterns in network traffic data \citep{berezinski2015entropy}. 
Specifically, for the sequential data we focus on in this paper, the extent of disorder in the data sets is reflected in the regularity of the order of nodes within the sequences. The block entropy (or n-gram entropy) is one of the most commonly used metrics to measure the extent of disorder in sequences \citep{schurmann1996entropy, jimenez2002entropy}. It is applicable when the number of distinct elements is relatively small and the probability of elements reappearing within the sequence is high, such as the DNA sequence composed of only 4 types of bases \citep{schmitt1997estimating}, and the song of humpback whales with a few dozens of acoustic signals \citep{suzuki2006information}. However, block entropy appears to struggle when faced with the information diffusion sequences due to its innate property (expounded in Section \ref{sec:apce}).

Quantifying the randomness of the data is often associated with the performance of the models. Some researchers have addressed the issue of the correlation between these two.
~\cite{song2010limits, lu2013approaching} use information entropy to quantify the ``predictability'' of human mobility, the upper bound that a person's next location can be predicted. ~\cite{chen2016temporal} apply a similar approach to investigate the ``predictability'' of users' online activities. ~\cite{lv2015toward, sun2020revealing} analyze the topological properties of the network and quantify the ``predictability'' toward the link prediction task. These works quantify the overall difficulties of the prediction. But the ``predictability'' is related to the theoretical bound that any method could ever achieve, which is not directly related to the performance of a particular prediction model. ~\cite{zhu2021sampling, zhan2022coarsas2hvec} use entropy-related measures to explain the performance of particular models, proving that the data quality sampled by the model is highly related to the model performance in tasks like node classification and link prediction. ~\cite{ran2024maximum} study the link prediction upper bound based on a particular topological feature and shows the relationship between the prediction accuracy and the topological characteristics of a network. ~\cite{barzel2013network} also demonstrate how the noise and uncertainty of data affect the performance of link prediction algorithm. However, due to these works only applying a limited number of measurements, one could not draw a consistent conclusion on how the performance varies with the changing complexity of the data. In this work, we identify a performance characteristic curve that illustrates a model's performance under different task complexity, mining the underlying pattern between data randomness and model performance comprehensively.

\section{Preliminary} \label{sec:pre}
\subsection{Definition of information diffusion prediction}

In this paper, the model evaluation task is conducted in the field of information diffusion prediction. The information diffusion prediction problem is described as follows. A social network is a graph $G=(V,E)$ in which each user in the network is the node $v \in V$ and each relation between users is the edge $e \in E$. When a user $v_0$ posts a message, other users in the network can forward it and these users compose an information diffusion sequence in which users are ordered by the forwarding time. The diffusion sequence which is also named a cascade is denoted as $c=(v_i|i=0,...,n)$ in which $v_0$ is the diffusion source. Then the information diffusion prediction problem is defined as given a cascade set $C=(c_k|k=1,...,m)$ consisting of $m$ cascades, to predict a cascade $\hat{c_p}$ for the ground truth $c_{p}=(v_i^{p}|i=0,...,n)$ when the new diffusion source $v_0^{p}$ of the cascade is known.

\subsection{Existing metric for prediction evaluation} \label{sec:evaluation_metric}        

As introduced in Section \ref{sec:related_studies}, there are several model evaluation metrics for the information diffusion prediction task, like RBP\citep{RBP}, HITS\citep{yang2019neural} and MAP\citep{MAP2008}. In this work, we choose MAP for its frequent usage in information diffusion prediction studies.

To calculate MAP, we need to first quantify the prediction accuracy of a single sequence. For a cascade sequence $c$ in the testing set (ground truth) and a predicted sequence $\hat{c}$, the prediction accuracy $\text{AP}_c$ is calculated as
\begin{equation}
AP_c=\frac{1}{|c|}\sum_{v \in c} \frac{|\hat{c}_{k(\hat{c},v)} \cap c|}{|\hat{c}_{k(\hat{c},v)}|},\label{eq:AP1}
\end{equation}
where $k(\hat{c},v)$ stands for the rank of node $v$ in the predicted cascade $\hat{c}$, $\hat{c}_{k(\hat{c},v)}$ stands for the top-$k(\hat{c},v)$ subsequence of $\hat{c}$. This subsequence is compared with $c$ and the overlap is then averaged to reach the final value of $\text{AP}_c$. The orders of nodes in the predicted and actual sequence are partially taken into consideration. If $c$ and $\hat{c}$ contain the same set of nodes, $\text{AP}_c = 1$ and nodes' order do not play a role. However, if nodes in $c$ and $\hat{c}$ are not the same, the match on the top part of the sequence is given a bigger weight.

The MAP is the average value of $\text{AP}_c$, calculated as
\begin{equation}
\text{MAP}=\frac{1}{\left|\mathcal{C}_{t}\right|} \sum_{c \in \mathcal{C}_{t}} AP_c, \label{eq:MAP3}
\end{equation}
where $\mathcal{C}_{t}$ is the set of predicted cascade sequences. 

\section{Performance characteristic curve for model evaluation} \label{sec:method}

\begin{figure*}[!htb]
	\centering
	\includegraphics[width=0.7\textwidth]{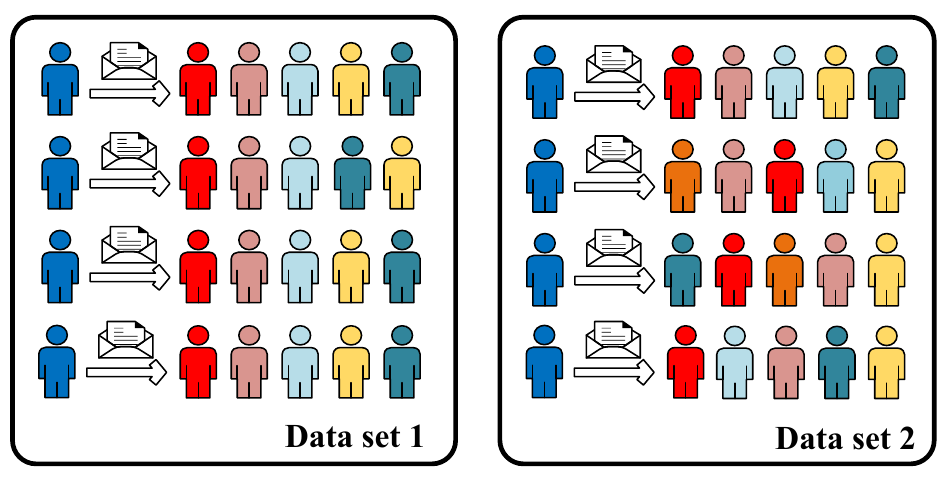}
	\caption{A simple example illustrating two information diffusion sequence data sets with definitely different extents of randomness and predictive difficulty. }\label{fig:apce_intro}
\end{figure*}

Existing metrics like MAP demonstrate the predictive ability of the model to some extent, but such metrics only focus on the performance of the model on a specific data set which is called single-point metrics. For example, as shown in Figure \ref{fig:apce_intro}, all models can achieve high scores on data set 1 because it is easy to predict obviously. However, a model that performs well on data set 2, which is highly random and disordered, is often what we need. Therefore, only considering the performance of the model on a single data set cannot provide a comprehensive evaluation of the model's predictive ability.

To address the shortcomings of single-point evaluation metrics like MAP, it is necessary to capture the comprehensive performance of the model facing different prediction scenarios. It naturally leads to the question of how to measure the randomness of data and relate it with the prediction accuracy of the model. In this section, we design an information-entropy-based metric to access the randomness in diffusion data and identify a scaling pattern between data randomness and the model's MAP on diffusion data sets, which provides a foundation for us to propose a more comprehensive model evaluation method.

\subsection{Metric for data randomness}\label{sec:apce}

As mentioned in Section \ref{sec:related_studies}, block entropy is commonly used to measure the degree of disorder in sequential data. We borrow the idea of block entropy and design a new metric to gauge the randomness of diffusion data in our work. The calculation of block entropy is designed as follows. 
Assume a sequence $[s_1,...s_t...,s_N]$. A window of length $n$ can be used to extract a subsequence (or block) like $[s_{t+1}, s_{t+2},...,s_{t+n}]$ from the original sequence, denoted by $x$. Through a sliding window, all subsequences are extracted from the sequence to form a set $X$, and the appearance frequency of each subsequence $x$ is counted, giving rise to $p(x)$. The block entropy is then calculated as
\begin{equation}
H_n = - \sum\limits_{x \in X}p_n(x)\log{p_n(x)}. \label{equation:blockentropy}
\end{equation}

In information spreading, the order that a message reaches different nodes is an important feature, reflecting properties such as the closeness to the source, the infection rates of individuals, the driven mechanism of the spreading, and more \citep{lee2010finding}. The consistency of the two nodes' relative positions reflects an extent of regularity in the data. Therefore, we turn to analyze the pairwise comparison of nodes' positions, leading to a $n \times n$ probability matrix $P=(p_{ij})_{i \not= j}$, where $p_{ij}$ is the probability that node $v_i$ ranks ahead of $v_j$. Note that $p_{ij} + p_{ji} = 1$ for all $i \not= j$ and $p_{ii}$ is undefined.

The probability $p_{ij}$ can be further applied to the formula of entropy, giving rise to a pairwise comparison entropy (PCE) as
\begin{equation}
\begin{array}{c}
\text{PCE}_{ij}= -\left(p_{ij} \log _{2} p_{ij}+{p}_{ji} \log _{2} {p}_{ji}\right). \label{eq:pair entropy}
\end{array}
\end{equation}
$\text{PCE}_{ij} = 0$ when the relative position of $v_i$ and $v_j$ is fixed in all sequences, indicating a high degree of regularity. On the contrary, $\text{PCE}_{ij}$ reaches the maximum when the relative position of node $v_i$ and $v_j$ are purely random.

By averaging the PCE of all node pairs, we obtain the average pairwise comparison entropy (APCE) associated with the overall extent of disorder
\begin{equation}
\text{APCE}=\sum_{(i,j) \in N} \frac{n_{i j}}{N} PCE_{ij}, \label{eq:APCE}
\end{equation}
where ${n_{i j}}$ is the number of times that node $v_i$ and $v_j$ simultaneously appear in the data, and $N$ is the total number of all node pairs. As an example, for two cascade sequence 
$$\left\{
\begin{array}{c}
1,2,3,4 \\
1,3,2
\end{array} \right\},$$
the APCE is calculated as 
\begin{equation}
\text{APCE} = (2{\text{PCE}_{12}}+2{\text{PCE}_{13}}+{\text{PCE}_{14}}+2{\text{PCE}_{23}}+{\text{PCE}_{24}}+{\text{PCE}_{34}} )/9= 2/9.\label{eq:apce_example}
\end{equation}

Note that different metrics are designed for different tasks. There is not an optimal metric that is applicable for all purposes. In this work, we aim to identify a universal relationship between the randomness of the data and the performance of a predictor. APCE works for this purpose. In contrast, when block entropy is applied, we could not get the pattern observed in the later part of the paper (Figure S4 in Supplementary Materials). Therefore, APCE is applied in this paper.

\subsection{Emergence of scaling pattern}

Using APCE to represent the randomness of diffusion data and MAP to represent the prediction performance of the model, we can investigate their relationship in a particular prediction model and establish a correlation pattern between APCE and MAP through extensive experiments.

\textbf{Model} We start with the content diffusion kernel (CDK) model, which uses a kernel motivated by heat diffusion to learn the latent representation of nodes \citep{CDK}. CDK is one of the earliest works that apply network embedding in information diffusion prediction. It relies on the temporal order of the spreading sequences, and does not require additional features such as the network structure or the user profile, which fits the scope of the problem studied here. Moreover, CDK has other variants with very similar designs, which can be used to validate the performance comparison. 

\textbf{Data} In order to obtain a sufficient number of diffusion sequence data to explore the pattern between sequence disorder and model accuracy, we collected massive sample sets from both synthetic and empirical data. For synthetic data, we generate the Erd\H{o}s-R\'enyi (ER) network \citep{erdHos1960evolution} and the Barab\'{a}si-Albert (BA) scale-free network by static model \citep{barabasi1999emergence} with a given average degree (ranges from 3 to 10) and network size (ranges from 100 to 1000 nodes). We run independent cascade (IC) mode \citep{kempe2003maximizing}, linear threshold (LT) model \citep{watts2002simple} and susceptible-infectious (SI) model \citep{hethcote2000mathematics} on these networks to generate synthetic spreading sequences. For SI and LT models, the simulation is carried out by the Gillespie algorithm which guarantees an accurate generation of the random process \citep{fennell2016limitations, ranyj}. The lower limit of the sequence length is set to 10 while the upper limit is 100. It needs to be emphasized that not all sequence lengths reach the upper limit. Based on the observation of empirical data, we limit the sequence length to 1/10 of the number of network nodes. For example, for a network consisting of 500 nodes, the sequence length we generated ranges from 10 to 50. In this way, we get 2640 synthetic sample sets in total.

For empirical data, we use spreading sequence on Twitter\citep{twt}, Digg\citep{digg}, and Douban, which records multiple lists of users sharing the same message ordered by time. The data set of Twitter contains 137,093 nodes (users in the social network), 3,589,811 edges (user links according to their following relationships), and 569 cascades. The data set of Digg contains 279,632 nodes, 2,617,993 edges and 3,553 cascades. The data set of Douban is collected by us, with the focus on the network centered on the ``Top 100 users'', which contains the 100 most popular users and their followers. The Douban data set contains 13,777 nodes, 567,250 edges, and 21,756 cascades.

The construction of empirical sample sets is similar to that on synthetic data. In these three empirical data sets, we select a fixed number of sequences from the entire set to form a sample set, and finally obtain 900 empirical diffusion sample sets (300 each).
To generate the spreading sequence with diverse complexity from empirical data, we select a fixed number of nodes from the original diffusion sequences while keeping the relative positions of selected nodes unchanged. Because the original sequence length in Twitter and Douban is barely more than 50, the length of the generated sequence in these two data sets ranges from 10 to 40. In Digg, the sequence length ranges from 10 to 100. 

\begin{figure}[!htb]
	\centering
	\includegraphics[width=1\textwidth]{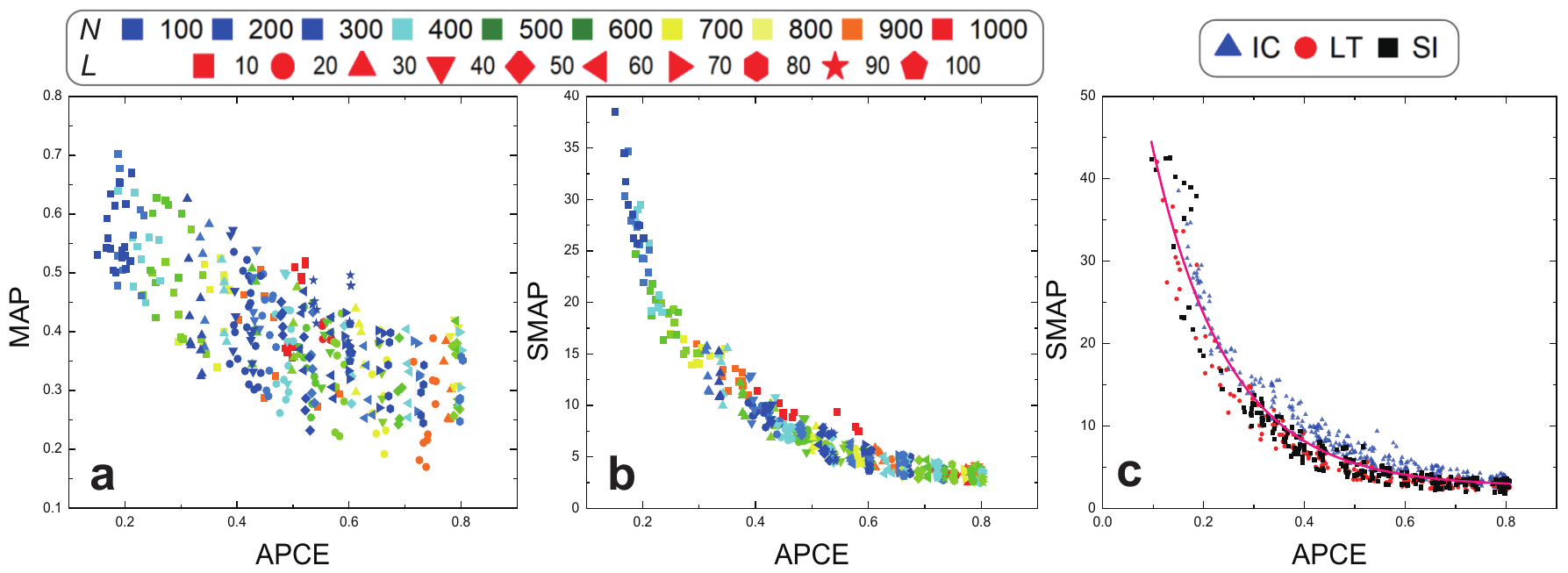}
	\caption{The correlation patterns of APCE and prediction performance. \textbf{a, b} The APCE-MAP pattern and APCE-SMAP pattern on cascade sets with various network sizes and cascade lengths generated by IC model. \textbf{c} The APCE-SMAP scaling pattern composed by ones on IC, LT and SI data sets. The goodness of the fitted curve of scaling pattern $R^2=0.94$. \label{fig:scaling_pattern}}
\end{figure}

\textbf{Experiment} To better identify potential patterns, we start with synthetic data that are less noisy. Experiments are firstly conducted on sample sets generated by IC model on ER networks. For a given sample set, we firstly employ the APCE metric to assess its degree of randomness. Subsequently, we utilize the CDK model to accomplish the task of diffusion prediction for this sample set. Finally, the prediction accuracy, evaluated by MAP, is obtained by comparing the predicted sequences with the actual sequences. Plotting the APCE and MAP values for all sample sets on a coordinate graph and the performance figure of the CDK model is obtained. However, we could not see a clear pattern (Figure \ref{fig:scaling_pattern}a). While MAP in general decreases with APCE as expected, data points are scattered. For data with a similar extent of the disorder, the prediction accuracy can fluctuate significantly. Hence, the direct relationship between APCE and MAP cannot generate a performance curve for the model, and therefore cannot accurately and comprehensively characterize the predictive ability of the model.

However, we notice that during the construction of diffusion sequence sample sets, the true factors fundamentally affecting the disorder of diffusion sequences are the topological characteristics of the network, such as network diameter, average path length, clustering coefficient \citep{newman2003structure}, and the characteristics of the diffusion process, including diffusion speed and branching degree \citep{leskovec2007patterns}. In contrast, parameters like the number of network nodes ($N$) and sequence length ($L$) are not inherently influential factors. Instead, they introduce noise into the model prediction and MAP (Mean Average Precision) calculations, thus impacting the observed patterns. The variation in the number of nodes affects model performance by altering the size of the prediction space, with larger node counts leading to lower model performance as there are more candidates to choose from. Meanwhile, longer sequence lengths increase the likelihood of higher precision in predicted sequences, thereby artificially inflating MAP values. The impact of $N$ and $L$ is related to the prediction task itself, not to the quality of the training data. Therefore, we should take them into account by resealing the original MAP, giving rise to the scaled MAP (SMAP) as
\begin{equation}
\text{SMAP} =  \text{MAP} \times \frac{N}{L}.
\end{equation}

When the accuracy metric is replaced by SMAP, we find that the originally scattered data points start to collapse on a single curve, demonstrating a clear scaling pattern (Figure \ref{fig:scaling_pattern}b). The same pattern also holds in synthetic data generated by LT and SI models (Figure S1 in Supplementary Materials). More importantly, when the data points from different spreading mechanisms are put together, they almost overlap each other, implying that the internally driven mechanism of the spreading has little impact on the scaling law (Figure \ref{fig:scaling_pattern}c). Instead, the network topology may be a more important factor. By changing the network topology from ER random network to BA scale-free network, we obtain the patterns that exhibit the same descending and scaling trend as those described in ER data, but the specific scaling curve differs slightly (Figure S2a in Supplementary Materials).

Given the nonlinear decay observed in the scaling curve, we apply an exponential decay function as $y = y_0 + Ae^{-Bx}$ for fitting. The curve is well fitted (Figure \ref{fig:scaling_pattern}c), capturing a model's inherent capability of making correct predictions against increased uncertainty. We regard it as the performance characteristic curve to evaluate the performance of information diffusion prediction models.

\section{Validation and application} \label{sec:experiments}
Does the scaling pattern observed between SMAP and APCE only hold for CDK model? Is the method of evaluating predictive ability using performance characteristic curves robust to other models? In this section, we address these two questions. We validate the feasibility of the proposed evaluation framework by using a family of models whose prediction abilities are theoretically known, serving as the ground truth.
Then the performance characteristic curve is applied to evaluate eight state-of-the-art models. In addition, we conduct a case study on models whose performance is hard to differentiate with conventional metrics, thereby demonstrating the superiority of our approach.

\subsection{Validation of performance characteristic curve}

We consider two variants of CDK that follow the general framework of CDK with slight modification of the choice embedding space. Instead of embedding all nodes in one latent space as CDK does, the position-aware asymmetric embedding (PAE) \citep{PAE} adopts an asymmetric embedding strategy to separately embed nodes into one influence space and one susceptibility space. The independent asymmetric embedding (IAE) \citep{IAE} embeds the source nodes into an influence space and embeds the infected nodes into $N$ susceptibility spaces, in order to avoid the mutual interference of embedding positions among the infected nodes of different cascades when they are in the same susceptibility latent space. %
The three models use the same heat diffusion kernel to learn the distance between nodes to predict the diffusion. Their only difference is in the number of latent space(s) used to represent a node. As the number of latent spaces increases, it is naturally expected, and also empirically tested that IAE would outperform PAE, and PAE would outperform CDK \citep{IAE}.

\begin{figure}[!htb]
	\centering
	\includegraphics[width=0.6\textwidth]{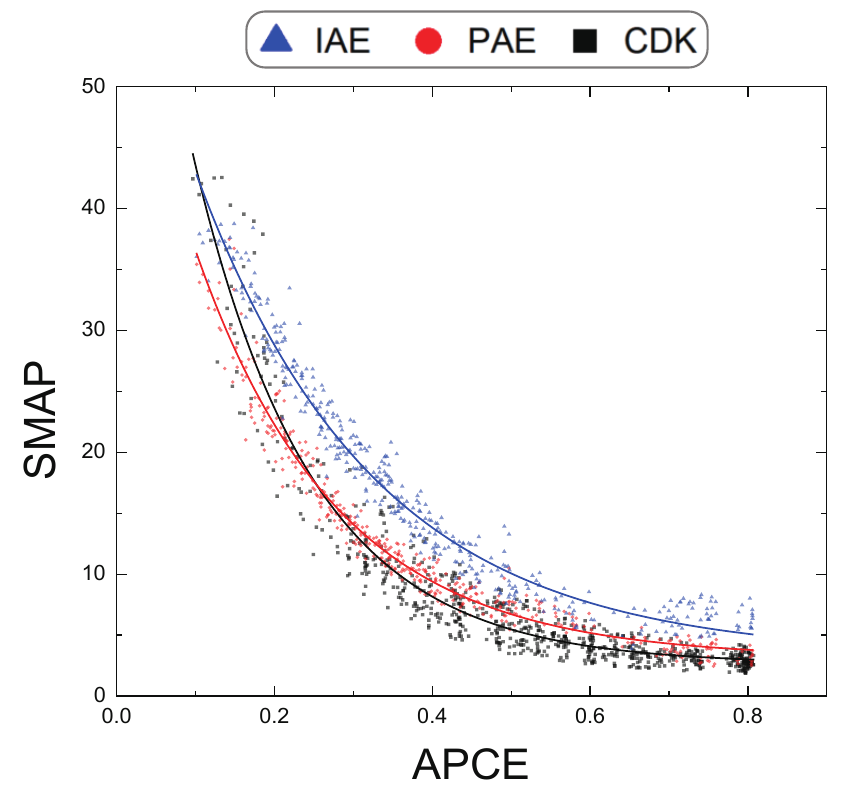}
	\caption{Scaling patterns of CDK, PAE, and IAE models on synthetic data. All the patterns of the three models are fitted well with the exponential decay functions. The goodness of fit are $R^{2}_{CDK}=0.94$, $R^{2}_{PAE}=0.97$, $R^{2}_{IAE}=0.97$. \label{fig:syn_3models}}
\end{figure}

As we have done with the CDK model, we test the PAE and IAE models on the same synthetic sample sets, ensuring that the training and testing sets remain unchanged. Since the prediction tasks are identical, the APCE values of the sample sets are consistent. Consequently, we can combine the scaling patterns of the three models to visually compare their performance on tasks of equivalent predictive difficulty (Figure \ref{fig:syn_3models}). This combined figure indicates that the scaling pattern holds in the other two models and the exponential decay function provides a good fit to the data points. Moreover, the fact that the scaling curves of each model are different from one to another confirms that the performance curve statistically captures a model's unique performance. Figure S2b in Supplementary Materials illustrates the similar scaling patterns of these three models on the synthetic BA data, which further proves the stability of the performance characteristic curve used for model evaluation, regardless of how the network topology changes.

Indeed, in Figure \ref{fig:syn_3models}, when the data is quite disordered (APCE is high), the curve of IAE is above the curve of PAE, and the curve of CDK is at the bottom. This confirms previous comparisons of performance among the three models \citep{PAE, IAE}. However, it is also interesting to note that when APCE is low, CDK's performance curve surpasses the other two. This suggests that when the data is less random and the prediction is consequently less challenging, using too many latent spaces can be overkill. 

We also investigate the SMAP-APCE plots in empirical data sets (Figure S3 in Supplementary Materials) and find similar behaviors. This similar comparison conclusion emerges in both synthetic and empirical data, showing the robustness of our evaluation method. Note that as a diffusion cascade set only involves a small proportion of the whole network nodes in empirical data, the network size $N$ used for scaling is set as the size of the diffusion subgraph.

\subsection{Application}
With the experiments that confirm the feasibility of utilizing the performance characteristics curve for model evaluation, we apply it to eight state-of-the-art models. We show that their performance can be well quantified and fairly compared.

\subsubsection{Evaluation of state-of-the-art models}
We select eight methods that can be applied to the information spreading task.

\textbf{Embedded-IC} \citep{bourigault2016representation} embeds users into a latent space based on the assumption that the underlying spreading mechanism follows the IC model. The relative positions of users in the latent space are then used to compute the diffusion probabilities between users.

\textbf{Topo-LSTM} \citep{wang2017topological} explores the diffusion topology of cascades using a directed acyclic graph. The graph is put into an LSTM-based model to generate topology-aware embeddings for users which are utilized for predicting nodes in a spreading.

\textbf{FOREST} \citep{yang2019neural} uses GRU to combine the temporal feature and user embedding. It employs the structural context extraction strategy to learn the underlying social graph, which is then sent to an MLP layer with the output of GRU to predict the diffusion probability.

\textbf{DCE} \citep{zhao2020deep} is an auto-encoder-based collaborative embedding model. It learns the node representations through cascade collaboration and node collaboration. Cascade collaboration captures the structural properties of nodes and the node collaboration captures the cascading context and cascading affinity.

\textbf{Dydiff-vae} \citep{wang2021dydiff} takes a dynamic encoder to infer the user interest in the next time step according to recent stimuli and social influence. It uses a dual attentive decoder to combine information content and user features. The original model takes into account the information content, which is unknown in the task discussed here. To apply this model, we use random vectors as the embedding of information content.

\textbf{MIDPMS} \citep{wang2023multiscale} models three types of features through the proposed minimal substitution neural network: information lifecycle, user preferences, and potential content expectations. It uses collaborative filtering to combine these features and predict information diffusion. Similar to the treatment of Dydiff-vae, random vectors are used as the embedding of information content.

\textbf{RotDiff} \citep{qiao2023rotdiff} uses the rotation transformation in the hyperbolic space to learn the embedding of nodes in both the social graph and diffusion graph. It uses the rotated Lorentz self-attention to extract the dependence of different diffusion sequences. Users' relative positions in the hyperbolic space are then used to compute the diffusion probabilities.

\textbf{STAHGCN} \citep{liu2023information} constructs a heterogeneous graph that combines user influence and user behavior. GCN is applied to find the graph embedding. The fusion mechanism is used to integrate node embeddings. Moreover, it utilizes time attention mechanism to encode the time feature.

\begin{figure}[!htb]
	\centering
	\includegraphics[width=1\textwidth]{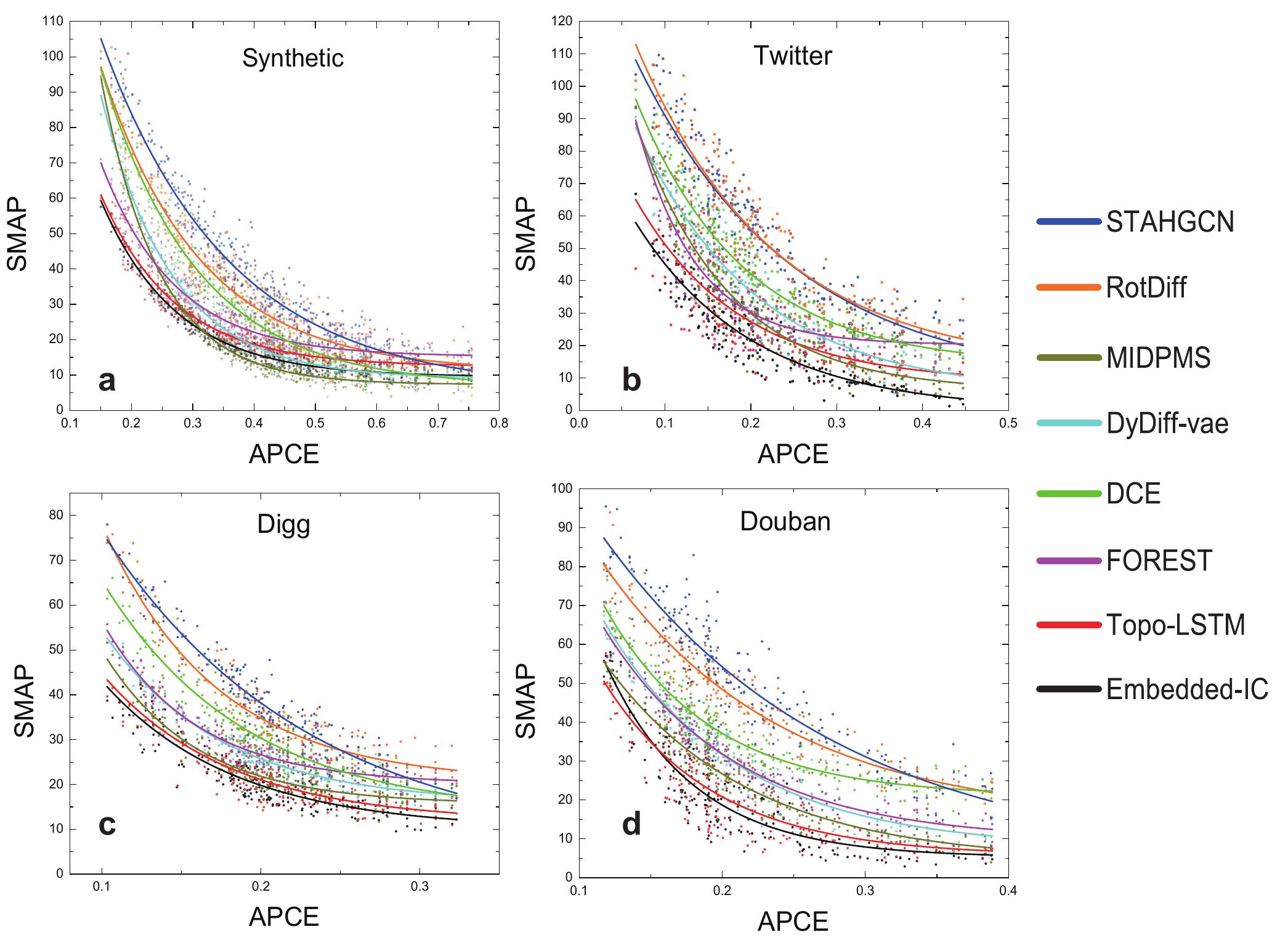}
	\caption{Application of performance characteristic curve for model evaluation on state-of-the-art models. \textbf{a} Models' APCE-SMAP patterns on synthetic data sets. \textbf{b, c, d} Models' APCE-SMAP patterns on Twitter, Digg, Douban data sets respectively.}\label{fig:sota_all}
\end{figure}

The model comparison based on the performance characteristic curve is illustrated in Figure \ref{fig:sota_all}. In general, RotDiff and STAHGCN, which are most recently introduced, exhibit superior performance compared to all others. RotDiff's performance is not as good as that of STAHGCN in regions when the task difficulty is not high. But RotDiff's performance drop is slower than that of STAHGCN. The performance curve predicts that RotDiff would surpass STAHGCN in very difficult prediction tasks. This may be attributed to RotDiff's utilization of hyperbolic space that well captures the asymmetrical characteristics within the diffusion process, making it more effective for tasks involving complex user influence dynamics and nonlinear diffusion pathways. The capabilities of DyDiff-vae and MIDPMS are compromised in this test, as the information content they incorporate is not available. MIDPMS is more negatively affected. The absence of information content makes the two models less effective than DCE that is proposed earlier in a wide range of task complexity. As expected, Embedded-IC and Topo-LSTM, the earliest in all eight models, are less effective than all others. The performance curve suggests that Topo-LSTM is better than Embedded-IC in most instances, but these two models' performances can be indistinguishable when the data is less random.

To summarize, the performance characteristic curve well captures the subtle differences among eight models. The conclusion is in line with our intuition as well as reported studies. More recent models tend to perform better than older ones. In addition, the curve also illustrates the dynamic changes in performance with different levels of task complexity. The best model in the low complexity region may become less effective as its performance drops faster than others.

\subsubsection{A case study}
In this section, we conduct a detailed case study on two pairs of models: Embedded-IC and Topo-LSTM, RotDiff and STAHGCN. In each pair, the model's capability is close to each other, which provides a representative example of the challenge in distinguishing model performance. To make the analysis more comprehensive, we take an additional performance measure HITS@k (Hit score on top-k) that is used in several recent studies \citep{qiao2023rotdiff, liu2023information, wang2023multiscale}. HITS@k is calculated as
\begin{equation}
\text{HITS@k}=\frac{1}{\left|\mathcal{C}_{t}\right|} \sum_{c \in \mathcal{C}_{t}} \frac{|\hat{c}_{k} \cap c_k|}{k}, \label{eq:hits}
\end{equation}
where $\mathcal{C}_{t}$ is the set of predicted cascade sequences; $\hat{c}_{k}$ and $c_{k}$ denote the top K nodes in the predicted sequence $\hat{c}$ and the ground truth sequence $c$, respectively. 
HITS@k quantifies the extent to which the nodes in the predicted sequences coincide with those in the ground truth sequences given a sequence length $k$. The performance of Embedded-IC, Topo-LSTM, RotDiff and STAHGCN are reported in Table \ref{tab:MAP_cs}, where MAP@k corresponds to the MAP values for predicted cascades with length $k$.

\begin{table}[htbp]
  \centering
  \caption{Two pairs of prediction models evaluated by metric MAP and HITS. The underline denotes the better performance between Embedded-IC and Topo-LSTM and the bold numbers indicate the better performance between RotDiff and STAHGCN.}
  \setstretch{1.2}
  \resizebox{\textwidth}{!}{
  \begin{threeparttable}
    \begin{tabular}{cc|ccccc|ccccc}
    \toprule
    \multirow{2}[4]{*}{Data set} & \multirow{2}[4]{*}{Method} & \multicolumn{5}{c|}{MAP@k(\%)}        & \multicolumn{5}{c}{HITS@k(\%)} \\
\cmidrule{3-12}          &       & @10   & @25   & @50   & @100  & @300  & @10   & @25   & @50   & @100  & @300 \\
    \midrule
    \multirow{4}[4]{*}{Twitter} & Embedded-IC & 0.203 & \underline{0.261} & \underline{0.514} & \underline{0.751} & 4.740  & 0.297 & 0.606 & \underline{1.169} & \underline{1.731} & \underline{8.379} \\
          & Topo-LSTM & \underline{0.215} & 0.243 & 0.509 & 0.665 & \underline{5.084} & \underline{0.323} & \underline{0.612} & 1.073 & 1.602 & 7.225 \\
\cmidrule{2-12}          & RotDiff & \textbf{0.256} & 0.462 & \textbf{1.435} & \textbf{2.857} & 14.532 & \textbf{0.382} & \textbf{1.626} & \textbf{2.893} & 4.752 & 25.383 \\
          & STAHGCN & 0.228 & \textbf{0.49} & 1.134 & 2.679 & \textbf{17.301} & 0.359 & 1.604 & 2.741 & \textbf{6.104} & \textbf{27.806} \\
    \midrule
    \multirow{4}[4]{*}{Digg} & Embedded-IC & \underline{0.327} & \underline{0.685} & 1.229 & 2.084 & 9.073 & 0.478 & 0.748 & 1.752 & 3.259 & 16.901 \\
          & Topo-LSTM & 0.304 & 0.679 & \underline{1.375} & \underline{2.444} & \underline{17.535} & \underline{0.562} & \underline{0.834} & \underline{2.576} & \underline{4.307} & \underline{22.835} \\
\cmidrule{2-12}          & RotDiff & \textbf{0.637} & 1.359 & 2.719 & 5.296 & \textbf{37.325} & \textbf{0.859} & 2.146 & 5.629 & 9.826 & 37.965 \\
          & STAHGCN & 0.529 & \textbf{1.824} & \textbf{3.382} & \textbf{6.581} & 35.908 & 0.825 & \textbf{2.758} & \textbf{6.713} & \textbf{12.147} & \textbf{44.327} \\
    \midrule
    \multirow{4}[4]{*}{Douban} & Embedded-IC & 0.228 & \underline{0.382} & 0.691 & \underline{1.216} & 3.251 & 0.476 & \underline{0.852} & \underline{1.711} & \underline{3.326} & 6.053 \\
          & Topo-LSTM & \underline{0.252} & 0.365 & \underline{0.784} & 0.935 & \underline{5.147} & \underline{0.483} & 0.698 & 1.467 & 2.107 & \underline{7.529} \\
\cmidrule{2-12}          & RotDiff & 0.329 & \textbf{1.258} & 2.116 & 3.517 & 18.182 & \textbf{0.589} & \textbf{1.764} & 3.271 & 5.183 & 28.594 \\
          & STAHGCN & \textbf{0.373} & 1.083 & \textbf{2.139} & \textbf{3.900} & \textbf{21.703} & 0.462 & 1.516 & \textbf{3.802} & \textbf{5.647} & \textbf{39.830} \\
    \bottomrule
    \end{tabular}%
      \end{threeparttable}}
  \label{tab:MAP_cs}%
\end{table}%

The static performance measure, either by MAP or HITS, demonstrates a certain extent of fluctuations. It is hard to draw a firm conclusion regarding which model is better. Indeed, by filtering the data or choosing the right hyper-parameters, one can claim that model A outperforms model B, or the opposite, revealing the limitation of static measure in model evaluation. On the contrary, the performance characteristic curve provides a model's performance within a specific range of task complexity, along with the changing dynamics in the face of escalating predictive challenges. This brings a more comprehensive model evaluation and selection. For instance, in the majority of scenarios, Topo-LSTM is the superior choice over Embedded-IC (Figure \ref{fig:sota_all}). However, when the task is less complicated (low APCE value of the data set), the two model's performance is comparable. Embedded-IC demonstrates a slight advantage on the Douban data. Taking into account the computational cost, the Embedded-IC becomes a recommended option in such scenario. Likewise, the performance characteristic curve suggests STAHGCN for the less challenging tasks and RotDiff for more difficult ones.

\section{Conclusion and future works} \label{sec:conclusion}

In this study, we aim to identify a performance characteristic curve for model evaluation and comparison. We focus specifically on the information diffusion prediction task. Traditionally, model evaluation takes only the static measure of the model performance in a particular data set. Here, we first assess the randomness across various cascade sets through an average pairwise comparison entropy, thereby reflecting the inherent complexity of the prediction task. 
Then we explore the model's performance across different levels of prediction complexity, and scale the accuracy measurement to a unified curve. This scaling curve epitomizes a model's inherent capability of making accurate predictions, serving as its performance characteristic curve. To the best of our knowledge, there is no similar approach that takes into consideration both the average performance under a specific task complexity and the dynamic changes.
Surpassing conventional approaches, this curve yields a visual depiction of the trade-off between data complexity and model accuracy, empowering users to make well-informed decisions in the selection and refinement of models for information diffusion prediction tasks.

The model evaluation approach presented in this paper is subject to limitations of computational cost. Calculating the APCE can be time-consuming, especially for long sequences. The generation of multiple spreading sequences with different APCE also requires additional processing time.
In addition, we acknowledge that this study only represents an initial step towards a more systematic and comprehensive evaluation of machine learning models. We select a relatively simple prediction task that relies solely on the sequence of user interactions, thereby eliminating other confounding factors that can be utilized for prediction. However, in domains such as computer vision and natural language processing, quantifying and modifying prediction complexity presents significant challenges. The performance curve measured in this work shows a network dependence. The network topology affects a model's performance curve. It can be interesting to investigate the way to include the topological feature and reach a more ``universal'' performance curve that holds in different networks. Finally, we note that the scaling pattern only holds for certain parameter regions that are mostly considered in prediction tasks. In extreme scenarios, the relationship between prediction accuracy and data randomness may diverge. Identifying the boundaries in which the proposed framework works can be an interesting problem for both theoretical and empirical studies.

\section*{CRediT authorship contribution statement}
\textbf{Wenjin Xie:} Conceptualization, Methodology, Software, Investigation, Visualization, Writing - original draft, Writing - review \& editing, Funding acquisition. \textbf{Xiaomeng Wang:} Conceptualization, Methodology, Writing - original draft, Funding acquisition. \textbf{Radosław Michalski:} Conceptualization, Writing – review \& editing, Funding acquisition. \textbf{Tao Jia:} Conceptualization, Methodology, Supervision, Formal analysis, Writing - review \& editing, Funding acquisition.

\section*{Data availability}
Data will be made available on request.

\section*{Acknowledgments}
This work is supported by the Chongqing Graduate Research and Innovation Project (No.CYB22128), the National Natural Science Foundation of China (NSFC) (No.62006198), the University Innovation Research Group of Chongqing (No. CXQT21005), the Fundamental Research Funds for the Central Universities (No. SWU-XDJH202303), the National Science Center, Poland (Grant No. 2021/41/B/HS6/02798), the Polish Ministry of Education and Science within the programme “International Projects Co-Funded”, and the European Union under the Horizon Europe (Grant No. 101086321 (OMINO)). However, the views and opinions expressed are those of the author(s) only and do not necessarily reflect those of the European Union or the European Research Executive Agency. Neither the European Union nor European Research Executive Agency can be held responsible for them.



\bibliography{reference}

\begin{thebibliography}{58}
\expandafter\ifx\csname natexlab\endcsname\relax\def\natexlab#1{#1}\fi
\providecommand{\url}[1]{\texttt{#1}}
\providecommand{\href}[2]{#2}
\providecommand{\path}[1]{#1}
\providecommand{\DOIprefix}{doi:}
\providecommand{\ArXivprefix}{arXiv:}
\providecommand{\URLprefix}{URL: }
\providecommand{\Pubmedprefix}{pmid:}
\providecommand{\doi}[1]{\href{http://dx.doi.org/#1}{\path{#1}}}
\providecommand{\Pubmed}[1]{\href{pmid:#1}{\path{#1}}}
\providecommand{\bibinfo}[2]{#2}
\ifx\xfnm\relax \def\xfnm[#1]{\unskip,\space#1}\fi
\bibitem[{Auxier \& Anderson(2021)}]{auxier2021social}
\bibinfo{author}{Auxier, B.}, \& \bibinfo{author}{Anderson, M.}
  (\bibinfo{year}{2021}).
\newblock \bibinfo{title}{Social media use in 2021}.
\newblock {\it \bibinfo{journal}{Pew Research Center}\/},  {\it
  \bibinfo{volume}{1}\/}, \bibinfo{pages}{1--4}.
\bibitem[{Barab{\'a}si \& Albert(1999)}]{barabasi1999emergence}
\bibinfo{author}{Barab{\'a}si, A.-L.}, \& \bibinfo{author}{Albert, R.}
  (\bibinfo{year}{1999}).
\newblock \bibinfo{title}{Emergence of scaling in random networks}.
\newblock {\it \bibinfo{journal}{science}\/},  {\it \bibinfo{volume}{286}\/},
  \bibinfo{pages}{509--512}.
\bibitem[{Barzel \& Barab{\'a}si(2013)}]{barzel2013network}
\bibinfo{author}{Barzel, B.}, \& \bibinfo{author}{Barab{\'a}si, A.-L.}
  (\bibinfo{year}{2013}).
\newblock \bibinfo{title}{Network link prediction by global silencing of
  indirect correlations}.
\newblock {\it \bibinfo{journal}{Nature biotechnology}\/},  {\it
  \bibinfo{volume}{31}\/}, \bibinfo{pages}{720--725}.
\bibitem[{Bengio et~al.(2009)Bengio, Louradour, Collobert \&
  Weston}]{bengio2009curriculum}
\bibinfo{author}{Bengio, Y.}, \bibinfo{author}{Louradour, J.},
  \bibinfo{author}{Collobert, R.}, \& \bibinfo{author}{Weston, J.}
  (\bibinfo{year}{2009}).
\newblock \bibinfo{title}{Curriculum learning}.
\newblock In {\it \bibinfo{booktitle}{Proceedings of the 26th annual
  international conference on machine learning}\/} (pp.
  \bibinfo{pages}{41--48}).
\bibitem[{Berezi{\'n}ski et~al.(2015)Berezi{\'n}ski, Jasiul \&
  Szpyrka}]{berezinski2015entropy}
\bibinfo{author}{Berezi{\'n}ski, P.}, \bibinfo{author}{Jasiul, B.}, \&
  \bibinfo{author}{Szpyrka, M.} (\bibinfo{year}{2015}).
\newblock \bibinfo{title}{An entropy-based network anomaly detection method}.
\newblock {\it \bibinfo{journal}{Entropy}\/},  {\it \bibinfo{volume}{17}\/},
  \bibinfo{pages}{2367--2408}.
\bibitem[{Berger et~al.(1996)Berger, Della~Pietra \&
  Della~Pietra}]{berger1996maximum}
\bibinfo{author}{Berger, A.}, \bibinfo{author}{Della~Pietra, S.~A.}, \&
  \bibinfo{author}{Della~Pietra, V.~J.} (\bibinfo{year}{1996}).
\newblock \bibinfo{title}{A maximum entropy approach to natural language
  processing}.
\newblock {\it \bibinfo{journal}{Computational linguistics}\/},  {\it
  \bibinfo{volume}{22}\/}, \bibinfo{pages}{39--71}.
\bibitem[{Bourigault et~al.(2014)Bourigault, Lagnier, Lamprier, Denoyer \&
  Gallinari}]{CDK}
\bibinfo{author}{Bourigault, S.}, \bibinfo{author}{Lagnier, C.},
  \bibinfo{author}{Lamprier, S.}, \bibinfo{author}{Denoyer, L.}, \&
  \bibinfo{author}{Gallinari, P.} (\bibinfo{year}{2014}).
\newblock \bibinfo{title}{Learning social network embeddings for predicting
  information diffusion}.
\newblock In {\it \bibinfo{booktitle}{Proceedings of the 7th ACM international
  conference on Web search and data mining}\/} (pp. \bibinfo{pages}{393--402}).
\newblock \bibinfo{organization}{ACM}.
\bibitem[{Bourigault et~al.(2016)Bourigault, Lamprier \&
  Gallinari}]{bourigault2016representation}
\bibinfo{author}{Bourigault, S.}, \bibinfo{author}{Lamprier, S.}, \&
  \bibinfo{author}{Gallinari, P.} (\bibinfo{year}{2016}).
\newblock \bibinfo{title}{Representation learning for information diffusion
  through social networks: an embedded cascade model}.
\newblock In {\it \bibinfo{booktitle}{Proceedings of the 9th ACM international
  conference on Web Search and Data Mining}\/} (pp. \bibinfo{pages}{573--582}).
\newblock \bibinfo{organization}{ACM}.
\bibitem[{Chen et~al.(2016)Chen, Gao \& Xiong}]{chen2016temporal}
\bibinfo{author}{Chen, W.}, \bibinfo{author}{Gao, Q.}, \&
  \bibinfo{author}{Xiong, H.} (\bibinfo{year}{2016}).
\newblock \bibinfo{title}{Temporal predictability of online behavior in
  foursquare}.
\newblock {\it \bibinfo{journal}{Entropy}\/},  {\it \bibinfo{volume}{18}\/},
  \bibinfo{pages}{296}.
\bibitem[{Erd{\H{o}}s \& R{\'e}nyi(1960)}]{erdHos1960evolution}
\bibinfo{author}{Erd{\H{o}}s, P.}, \& \bibinfo{author}{R{\'e}nyi, A.}
  (\bibinfo{year}{1960}).
\newblock \bibinfo{title}{On the evolution of random graphs}.
\newblock {\it \bibinfo{journal}{Publ. Math. Inst. Hung. Acad. Sci}\/},  {\it
  \bibinfo{volume}{5}\/}, \bibinfo{pages}{17--60}.
\bibitem[{Fennell et~al.(2016)Fennell, Melnik \&
  Gleeson}]{fennell2016limitations}
\bibinfo{author}{Fennell, P.~G.}, \bibinfo{author}{Melnik, S.}, \&
  \bibinfo{author}{Gleeson, J.~P.} (\bibinfo{year}{2016}).
\newblock \bibinfo{title}{Limitations of discrete-time approaches to
  continuous-time contagion dynamics}.
\newblock {\it \bibinfo{journal}{Physical Review E}\/},  {\it
  \bibinfo{volume}{94}\/}, \bibinfo{pages}{052125}.
\bibitem[{Hethcote(1989)}]{hethcote1989three}
\bibinfo{author}{Hethcote, H.~W.} (\bibinfo{year}{1989}).
\newblock \bibinfo{title}{Three basic epidemiological models}.
\newblock In {\it \bibinfo{booktitle}{Proceedings of Applied mathematical
  ecology}\/} (pp. \bibinfo{pages}{119--144}).
\newblock \bibinfo{publisher}{Springer}.
\bibitem[{Hethcote(2000)}]{hethcote2000mathematics}
\bibinfo{author}{Hethcote, H.~W.} (\bibinfo{year}{2000}).
\newblock \bibinfo{title}{The mathematics of infectious diseases}.
\newblock {\it \bibinfo{journal}{SIAM review}\/},  {\it
  \bibinfo{volume}{42}\/}, \bibinfo{pages}{599--653}.
\bibitem[{Hodas \& Lerman(2014)}]{twt}
\bibinfo{author}{Hodas, N.~O.}, \& \bibinfo{author}{Lerman, K.}
  (\bibinfo{year}{2014}).
\newblock \bibinfo{title}{The simple rules of social contagion}.
\newblock {\it \bibinfo{journal}{Scientific reports}\/},  {\it
  \bibinfo{volume}{4}\/}, \bibinfo{pages}{4343}.
\bibitem[{Hofman et~al.(2017)Hofman, Sharma \& Watts}]{hofman2017prediction}
\bibinfo{author}{Hofman, J.~M.}, \bibinfo{author}{Sharma, A.}, \&
  \bibinfo{author}{Watts, D.~J.} (\bibinfo{year}{2017}).
\newblock \bibinfo{title}{Prediction and explanation in social systems}.
\newblock {\it \bibinfo{journal}{Science}\/},  {\it \bibinfo{volume}{355}\/},
  \bibinfo{pages}{486--488}.
\bibitem[{Hogg \& Lerman(2012)}]{digg}
\bibinfo{author}{Hogg, T.}, \& \bibinfo{author}{Lerman, K.}
  (\bibinfo{year}{2012}).
\newblock \bibinfo{title}{Social dynamics of digg}.
\newblock {\it \bibinfo{journal}{EPJ Data Science}\/},  {\it
  \bibinfo{volume}{1}\/}, \bibinfo{pages}{5}.
\bibitem[{Jim{\'e}nez-Monta{\~n}o et~al.(2002)Jim{\'e}nez-Monta{\~n}o, Ebeling,
  Pohl \& Rapp}]{jimenez2002entropy}
\bibinfo{author}{Jim{\'e}nez-Monta{\~n}o, M.~A.}, \bibinfo{author}{Ebeling,
  W.}, \bibinfo{author}{Pohl, T.}, \& \bibinfo{author}{Rapp, P.~E.}
  (\bibinfo{year}{2002}).
\newblock \bibinfo{title}{Entropy and complexity of finite sequences as
  fluctuating quantities}.
\newblock {\it \bibinfo{journal}{Biosystems}\/},  {\it \bibinfo{volume}{64}\/},
  \bibinfo{pages}{23--32}.
\bibitem[{Kempe et~al.(2003)Kempe, Kleinberg \& Tardos}]{kempe2003maximizing}
\bibinfo{author}{Kempe, D.}, \bibinfo{author}{Kleinberg, J.}, \&
  \bibinfo{author}{Tardos, {\'E}.} (\bibinfo{year}{2003}).
\newblock \bibinfo{title}{Maximizing the spread of influence through a social
  network}.
\newblock In {\it \bibinfo{booktitle}{Proceedings of the 9th ACM SIGKDD
  international conference on Knowledge discovery and data mining}\/} (pp.
  \bibinfo{pages}{137--146}).
\newblock \bibinfo{organization}{ACM}.
\bibitem[{Kunaver \& Po{\v{z}}rl(2017)}]{kunaver2017diversity}
\bibinfo{author}{Kunaver, M.}, \& \bibinfo{author}{Po{\v{z}}rl, T.}
  (\bibinfo{year}{2017}).
\newblock \bibinfo{title}{Diversity in recommender systems--a survey}.
\newblock {\it \bibinfo{journal}{Knowledge-based systems}\/},  {\it
  \bibinfo{volume}{123}\/}, \bibinfo{pages}{154--162}.
\bibitem[{Lee et~al.(2010)Lee, Kwak, Park \& Moon}]{lee2010finding}
\bibinfo{author}{Lee, C.}, \bibinfo{author}{Kwak, H.}, \bibinfo{author}{Park,
  H.}, \& \bibinfo{author}{Moon, S.} (\bibinfo{year}{2010}).
\newblock \bibinfo{title}{Finding influentials based on the temporal order of
  information adoption in twitter}.
\newblock In {\it \bibinfo{booktitle}{Proceedings of the 19th international
  conference on World Wide Web}\/} (pp. \bibinfo{pages}{1137--1138}).
\bibitem[{Leskovec et~al.(2007)Leskovec, McGlohon, Faloutsos, Glance \&
  Hurst}]{leskovec2007patterns}
\bibinfo{author}{Leskovec, J.}, \bibinfo{author}{McGlohon, M.},
  \bibinfo{author}{Faloutsos, C.}, \bibinfo{author}{Glance, N.}, \&
  \bibinfo{author}{Hurst, M.} (\bibinfo{year}{2007}).
\newblock \bibinfo{title}{Patterns of cascading behavior in large blog graphs}.
\newblock In {\it \bibinfo{booktitle}{Proceedings of the 2007 SIAM
  international conference on data mining}\/} (pp. \bibinfo{pages}{551--556}).
\newblock \bibinfo{organization}{SIAM}.
\bibitem[{Liu et~al.(2014)Liu, Liu, Huang \& Bovik}]{liu2014no}
\bibinfo{author}{Liu, L.}, \bibinfo{author}{Liu, B.}, \bibinfo{author}{Huang,
  H.}, \& \bibinfo{author}{Bovik, A.~C.} (\bibinfo{year}{2014}).
\newblock \bibinfo{title}{No-reference image quality assessment based on
  spatial and spectral entropies}.
\newblock {\it \bibinfo{journal}{Signal processing: Image communication}\/},
  {\it \bibinfo{volume}{29}\/}, \bibinfo{pages}{856--863}.
\bibitem[{Liu et~al.(2016)Liu, Shen, Ouyang, Fu, Zha \& Cheng}]{PAE}
\bibinfo{author}{Liu, W.}, \bibinfo{author}{Shen, H.}, \bibinfo{author}{Ouyang,
  W.}, \bibinfo{author}{Fu, G.}, \bibinfo{author}{Zha, L.}, \&
  \bibinfo{author}{Cheng, X.} (\bibinfo{year}{2016}).
\newblock \bibinfo{title}{Learning cost-effective social embedding for cascade
  prediction}.
\newblock In {\it \bibinfo{booktitle}{Proceedings of the Chinese National
  Conference on Social Media Processing}\/} (pp. \bibinfo{pages}{1--13}).
\newblock \bibinfo{organization}{Springer}.
\bibitem[{Liu et~al.(2023)Liu, Miao, Fiumara \& De~Meo}]{liu2023information}
\bibinfo{author}{Liu, X.}, \bibinfo{author}{Miao, C.},
  \bibinfo{author}{Fiumara, G.}, \& \bibinfo{author}{De~Meo, P.}
  (\bibinfo{year}{2023}).
\newblock \bibinfo{title}{Information propagation prediction based on
  spatial--temporal attention and heterogeneous graph convolutional networks}.
\newblock {\it \bibinfo{journal}{IEEE Transactions on Computational Social
  Systems}\/}, .
\bibitem[{Lu et~al.(2013)Lu, Wetter, Bharti, Tatem \&
  Bengtsson}]{lu2013approaching}
\bibinfo{author}{Lu, X.}, \bibinfo{author}{Wetter, E.},
  \bibinfo{author}{Bharti, N.}, \bibinfo{author}{Tatem, A.~J.}, \&
  \bibinfo{author}{Bengtsson, L.} (\bibinfo{year}{2013}).
\newblock \bibinfo{title}{Approaching the limit of predictability in human
  mobility}.
\newblock {\it \bibinfo{journal}{Scientific Reports}\/},  {\it
  \bibinfo{volume}{3}\/}, \bibinfo{pages}{2923--2923}.
\bibitem[{Lü et~al.(2015)Lü, Pan, Zhou, Zhang \& Stanley}]{lv2015toward}
\bibinfo{author}{Lü, L.}, \bibinfo{author}{Pan, L.}, \bibinfo{author}{Zhou,
  T.}, \bibinfo{author}{Zhang, Y.-C.}, \& \bibinfo{author}{Stanley, H.~E.}
  (\bibinfo{year}{2015}).
\newblock \bibinfo{title}{Toward link predictability of complex networks}.
\newblock {\it \bibinfo{journal}{Proceedings of the National Academy of
  Sciences of the United States of America}\/},  {\it \bibinfo{volume}{112}\/},
  \bibinfo{pages}{2325--2330}.
\bibitem[{MacKay(2003)}]{mackay2003information}
\bibinfo{author}{MacKay, D.~J.} (\bibinfo{year}{2003}).
\newblock {\it \bibinfo{title}{Information theory, inference and learning
  algorithms}\/}.
\newblock \bibinfo{publisher}{Cambridge university press}.
\bibitem[{Martin et~al.(2016)Martin, Hofman, Sharma, Anderson \&
  Watts}]{martin2016exploring}
\bibinfo{author}{Martin, T.}, \bibinfo{author}{Hofman, J.~M.},
  \bibinfo{author}{Sharma, A.}, \bibinfo{author}{Anderson, A.}, \&
  \bibinfo{author}{Watts, D.~J.} (\bibinfo{year}{2016}).
\newblock \bibinfo{title}{Exploring limits to prediction in complex social
  systems}.
\newblock In {\it \bibinfo{booktitle}{Proceedings of the 25th international
  conference on World Wide Web}\/} (pp. \bibinfo{pages}{683--694}).
\bibitem[{Michalski et~al.(2022)Michalski, Serwata, Nurek, Szymanski, Kazienko
  \& Jia}]{michalski2022temporal}
\bibinfo{author}{Michalski, R.}, \bibinfo{author}{Serwata, D.},
  \bibinfo{author}{Nurek, M.}, \bibinfo{author}{Szymanski, B.~K.},
  \bibinfo{author}{Kazienko, P.}, \& \bibinfo{author}{Jia, T.}
  (\bibinfo{year}{2022}).
\newblock \bibinfo{title}{Temporal network epistemology: On reaching consensus
  in a real-world setting}.
\newblock {\it \bibinfo{journal}{Chaos: An Interdisciplinary Journal of
  Nonlinear Science}\/},  {\it \bibinfo{volume}{32}\/},
  \bibinfo{pages}{063135}.
\bibitem[{Moffat et~al.(2007)Moffat, Webber \& Zobel}]{RBP}
\bibinfo{author}{Moffat, A.}, \bibinfo{author}{Webber, W.}, \&
  \bibinfo{author}{Zobel, J.} (\bibinfo{year}{2007}).
\newblock \bibinfo{title}{Strategic system comparisons via targeted relevance
  judgments}.
\newblock In {\it \bibinfo{booktitle}{Proceedings of the 30th annual
  international ACM SIGIR conference on Research and development in information
  retrieval}\/} (pp. \bibinfo{pages}{375--382}).
\bibitem[{Newman(2003)}]{newman2003structure}
\bibinfo{author}{Newman, M.~E.} (\bibinfo{year}{2003}).
\newblock \bibinfo{title}{The structure and function of complex networks}.
\newblock {\it \bibinfo{journal}{SIAM review}\/},  {\it
  \bibinfo{volume}{45}\/}, \bibinfo{pages}{167--256}.
\bibitem[{Pentina \& Lampert(2014)}]{pentina2014pac}
\bibinfo{author}{Pentina, A.}, \& \bibinfo{author}{Lampert, C.}
  (\bibinfo{year}{2014}).
\newblock \bibinfo{title}{A pac-bayesian bound for lifelong learning}.
\newblock In {\it \bibinfo{booktitle}{International Conference on Machine
  Learning}\/} (pp. \bibinfo{pages}{991--999}).
\newblock \bibinfo{organization}{PMLR}.
\bibitem[{Qiao et~al.(2023)Qiao, Feng, Li, Lin, Hu, Wei \&
  Ye}]{qiao2023rotdiff}
\bibinfo{author}{Qiao, H.}, \bibinfo{author}{Feng, S.}, \bibinfo{author}{Li,
  X.}, \bibinfo{author}{Lin, H.}, \bibinfo{author}{Hu, H.},
  \bibinfo{author}{Wei, W.}, \& \bibinfo{author}{Ye, Y.}
  (\bibinfo{year}{2023}).
\newblock \bibinfo{title}{Rotdiff: A hyperbolic rotation representation model
  for information diffusion prediction}.
\newblock In {\it \bibinfo{booktitle}{Proceedings of the 32nd ACM International
  Conference on Information and Knowledge Management}\/} (pp.
  \bibinfo{pages}{2065--2074}).
\bibitem[{Rainio et~al.(2024)Rainio, Teuho \& Kl{\'e}n}]{rainio2024evaluation}
\bibinfo{author}{Rainio, O.}, \bibinfo{author}{Teuho, J.}, \&
  \bibinfo{author}{Kl{\'e}n, R.} (\bibinfo{year}{2024}).
\newblock \bibinfo{title}{Evaluation metrics and statistical tests for machine
  learning}.
\newblock {\it \bibinfo{journal}{Scientific Reports}\/},  {\it
  \bibinfo{volume}{14}\/}, \bibinfo{pages}{6086}.
\bibitem[{Ran et~al.(2020)Ran, Deng, Wang \& Jia}]{ranyj}
\bibinfo{author}{Ran, Y.}, \bibinfo{author}{Deng, X.}, \bibinfo{author}{Wang,
  X.}, \& \bibinfo{author}{Jia, T.} (\bibinfo{year}{2020}).
\newblock \bibinfo{title}{A generalized linear threshold model for an improved
  description of the spreading dynamics}.
\newblock {\it \bibinfo{journal}{Chaos: An Interdisciplinary Journal of
  Nonlinear Science}\/},  {\it \bibinfo{volume}{30}\/},
  \bibinfo{pages}{083127}.
\bibitem[{Ran et~al.(2024)Ran, Xu \& Jia}]{ran2024maximum}
\bibinfo{author}{Ran, Y.}, \bibinfo{author}{Xu, X.-K.}, \&
  \bibinfo{author}{Jia, T.} (\bibinfo{year}{2024}).
\newblock \bibinfo{title}{The maximum capability of a topological feature in
  link prediction}.
\newblock {\it \bibinfo{journal}{PNAS nexus}\/},  {\it \bibinfo{volume}{3}\/},
  \bibinfo{pages}{pgae113}.
\bibitem[{Raschka(2018)}]{raschka2018model}
\bibinfo{author}{Raschka, S.} (\bibinfo{year}{2018}).
\newblock \bibinfo{title}{Model evaluation, model selection, and algorithm
  selection in machine learning}.
\newblock {\it \bibinfo{journal}{arXiv preprint arXiv:1811.12808}\/}, .
\bibitem[{Robertson(2008)}]{MAP2008}
\bibinfo{author}{Robertson, S.} (\bibinfo{year}{2008}).
\newblock \bibinfo{title}{A new interpretation of average precision}.
\newblock In {\it \bibinfo{booktitle}{Proceedings of the 31st annual
  international ACM SIGIR conference on Research and development in information
  retrieval}\/} (pp. \bibinfo{pages}{689--690}).
\bibitem[{Schelter et~al.(2018)Schelter, Biessmann, Januschowski, Salinas,
  Seufert \& Szarvas}]{Schelter2015}
\bibinfo{author}{Schelter, S.}, \bibinfo{author}{Biessmann, F.},
  \bibinfo{author}{Januschowski, T.}, \bibinfo{author}{Salinas, D.},
  \bibinfo{author}{Seufert, S.}, \& \bibinfo{author}{Szarvas, G.}
  (\bibinfo{year}{2018}).
\newblock \bibinfo{title}{On challenges in machine learning model management}.
\newblock {\it \bibinfo{journal}{IEEE Data Engineering Bulletin}\/},  {\it
  \bibinfo{volume}{41}\/}, \bibinfo{pages}{5--15}.
\bibitem[{Schmitt \& Herzel(1997)}]{schmitt1997estimating}
\bibinfo{author}{Schmitt, A.~O.}, \& \bibinfo{author}{Herzel, H.}
  (\bibinfo{year}{1997}).
\newblock \bibinfo{title}{Estimating the entropy of dna sequences}.
\newblock {\it \bibinfo{journal}{Journal of theoretical biology}\/},  {\it
  \bibinfo{volume}{188}\/}, \bibinfo{pages}{369--377}.
\bibitem[{Sch{\"u}rmann \& Grassberger(1996)}]{schurmann1996entropy}
\bibinfo{author}{Sch{\"u}rmann, T.}, \& \bibinfo{author}{Grassberger, P.}
  (\bibinfo{year}{1996}).
\newblock \bibinfo{title}{Entropy estimation of symbol sequences}.
\newblock {\it \bibinfo{journal}{Chaos: An Interdisciplinary Journal of
  Nonlinear Science}\/},  {\it \bibinfo{volume}{6}\/},
  \bibinfo{pages}{414--427}.
\bibitem[{Shannon(1948)}]{shannon1948mathematical}
\bibinfo{author}{Shannon, C.~E.} (\bibinfo{year}{1948}).
\newblock \bibinfo{title}{A mathematical theory of communication}.
\newblock {\it \bibinfo{journal}{Bell system technical journal}\/},  {\it
  \bibinfo{volume}{27}\/}, \bibinfo{pages}{379--423}.
\bibitem[{Song et~al.(2010)Song, Qu, Blumm \& Barabási}]{song2010limits}
\bibinfo{author}{Song, C.}, \bibinfo{author}{Qu, Z.}, \bibinfo{author}{Blumm,
  N.}, \& \bibinfo{author}{Barabási, A.~L.} (\bibinfo{year}{2010}).
\newblock \bibinfo{title}{Limits of predictability in human mobility}.
\newblock {\it \bibinfo{journal}{Science}\/},  {\it \bibinfo{volume}{327}\/},
  \bibinfo{pages}{1018--1021}.
\bibitem[{Su(1992)}]{su1992evaluation}
\bibinfo{author}{Su, L.~T.} (\bibinfo{year}{1992}).
\newblock \bibinfo{title}{Evaluation measures for interactive information
  retrieval}.
\newblock {\it \bibinfo{journal}{Information Processing \& Management}\/},
  {\it \bibinfo{volume}{28}\/}, \bibinfo{pages}{503--516}.
\bibitem[{Sun et~al.(2020)Sun, Feng, Xie, Ma, Wang \& Hu}]{sun2020revealing}
\bibinfo{author}{Sun, J.}, \bibinfo{author}{Feng, L.}, \bibinfo{author}{Xie,
  J.}, \bibinfo{author}{Ma, X.}, \bibinfo{author}{Wang, D.}, \&
  \bibinfo{author}{Hu, Y.} (\bibinfo{year}{2020}).
\newblock \bibinfo{title}{Revealing the predictability of intrinsic structure
  in complex networks}.
\newblock {\it \bibinfo{journal}{Nature communications}\/},  {\it
  \bibinfo{volume}{11}\/}, \bibinfo{pages}{1--10}.
\bibitem[{Suzuki et~al.(2006)Suzuki, Buck \& Tyack}]{suzuki2006information}
\bibinfo{author}{Suzuki, R.}, \bibinfo{author}{Buck, J.~R.}, \&
  \bibinfo{author}{Tyack, P.~L.} (\bibinfo{year}{2006}).
\newblock \bibinfo{title}{Information entropy of humpback whale songs}.
\newblock {\it \bibinfo{journal}{The Journal of the Acoustical Society of
  America}\/},  {\it \bibinfo{volume}{119}\/}, \bibinfo{pages}{1849--1866}.
\bibitem[{Wang et~al.(2017)Wang, Zheng, Liu \& Chang}]{wang2017topological}
\bibinfo{author}{Wang, J.}, \bibinfo{author}{Zheng, V.~W.},
  \bibinfo{author}{Liu, Z.}, \& \bibinfo{author}{Chang, K. C.-C.}
  (\bibinfo{year}{2017}).
\newblock \bibinfo{title}{Topological recurrent neural network for diffusion
  prediction}.
\newblock In {\it \bibinfo{booktitle}{Proceedings of the International
  Conference on Data Mining}\/} (pp. \bibinfo{pages}{475--484}).
\newblock \bibinfo{organization}{IEEE}.
\bibitem[{Wang et~al.(2021)Wang, Huang, Liu, Shao, Liu, Li, Wang, Sun, Yao \&
  Abdelzaher}]{wang2021dydiff}
\bibinfo{author}{Wang, R.}, \bibinfo{author}{Huang, Z.}, \bibinfo{author}{Liu,
  S.}, \bibinfo{author}{Shao, H.}, \bibinfo{author}{Liu, D.},
  \bibinfo{author}{Li, J.}, \bibinfo{author}{Wang, T.}, \bibinfo{author}{Sun,
  D.}, \bibinfo{author}{Yao, S.}, \& \bibinfo{author}{Abdelzaher, T.}
  (\bibinfo{year}{2021}).
\newblock \bibinfo{title}{Dydiff-vae: A dynamic variational framework for
  information diffusion prediction}.
\newblock In {\it \bibinfo{booktitle}{Proceedings of the 44th International ACM
  SIGIR Conference on Research and Development in Information Retrieval}\/}
  (pp. \bibinfo{pages}{163--172}).
\bibitem[{Wang et~al.(2023)Wang, Xu \& Zhang}]{wang2023multiscale}
\bibinfo{author}{Wang, R.}, \bibinfo{author}{Xu, X.}, \&
  \bibinfo{author}{Zhang, Y.} (\bibinfo{year}{2023}).
\newblock \bibinfo{title}{Multiscale information diffusion prediction with
  minimal substitution neural network.}
\newblock {\it \bibinfo{journal}{IEEE Transactions on Neural Networks and
  Learning Systems}\/}, .
\bibitem[{Watts(2002)}]{watts2002simple}
\bibinfo{author}{Watts, D.~J.} (\bibinfo{year}{2002}).
\newblock \bibinfo{title}{A simple model of global cascades on random
  networks}.
\newblock {\it \bibinfo{journal}{Proceedings of the National Academy of
  Sciences}\/},  {\it \bibinfo{volume}{99}\/}, \bibinfo{pages}{5766--5771}.
\bibitem[{Xie et~al.(2021)Xie, Meng, Sun, Ma, Yan \& Hu}]{xie2021detecting}
\bibinfo{author}{Xie, J.}, \bibinfo{author}{Meng, F.}, \bibinfo{author}{Sun,
  J.}, \bibinfo{author}{Ma, X.}, \bibinfo{author}{Yan, G.}, \&
  \bibinfo{author}{Hu, Y.} (\bibinfo{year}{2021}).
\newblock \bibinfo{title}{Detecting and modelling real percolation and phase
  transitions of information on social media}.
\newblock {\it \bibinfo{journal}{Nature Human Behaviour}\/},  {\it
  \bibinfo{volume}{5}\/}, \bibinfo{pages}{1161--1168}.
\bibitem[{Xie et~al.(2022)Xie, Wang \& Jia}]{IAE}
\bibinfo{author}{Xie, W.}, \bibinfo{author}{Wang, X.}, \& \bibinfo{author}{Jia,
  T.} (\bibinfo{year}{2022}).
\newblock \bibinfo{title}{Independent asymmetric embedding for information
  diffusion prediction on social networks}.
\newblock In {\it \bibinfo{booktitle}{Proceedings of the 25th International
  Conference on Computer Supported Cooperative Work in Design}\/} (pp.
  \bibinfo{pages}{190--195}).
\newblock \bibinfo{organization}{IEEE}.
\bibitem[{Yang et~al.(2019)Yang, Sun, Liu, Han, Liu \& Luan}]{yang2019neural}
\bibinfo{author}{Yang, C.}, \bibinfo{author}{Sun, M.}, \bibinfo{author}{Liu,
  H.}, \bibinfo{author}{Han, S.}, \bibinfo{author}{Liu, Z.}, \&
  \bibinfo{author}{Luan, H.} (\bibinfo{year}{2019}).
\newblock \bibinfo{title}{Neural diffusion model for microscopic cascade
  study}.
\newblock {\it \bibinfo{journal}{IEEE Transactions on Knowledge and Data
  Engineering}\/},  {\it \bibinfo{volume}{33}\/}, \bibinfo{pages}{1128--1139}.
\bibitem[{Zamir et~al.(2018)Zamir, Sax, Shen, Guibas, Malik \&
  Savarese}]{zamir2018taskonomy}
\bibinfo{author}{Zamir, A.~R.}, \bibinfo{author}{Sax, A.},
  \bibinfo{author}{Shen, W.}, \bibinfo{author}{Guibas, L.~J.},
  \bibinfo{author}{Malik, J.}, \& \bibinfo{author}{Savarese, S.}
  (\bibinfo{year}{2018}).
\newblock \bibinfo{title}{Taskonomy: Disentangling task transfer learning}.
\newblock In {\it \bibinfo{booktitle}{Proceedings of the IEEE conference on
  computer vision and pattern recognition}\/} (pp.
  \bibinfo{pages}{3712--3722}).
\bibitem[{Zhan \& Jia(2022)}]{zhan2022coarsas2hvec}
\bibinfo{author}{Zhan, L.}, \& \bibinfo{author}{Jia, T.}
  (\bibinfo{year}{2022}).
\newblock \bibinfo{title}{Coarsas2hvec: Heterogeneous information network
  embedding with balanced network sampling}.
\newblock {\it \bibinfo{journal}{Entropy}\/},  {\it \bibinfo{volume}{24}\/},
  \bibinfo{pages}{276}.
\bibitem[{Zhang et~al.(2022)Zhang, Guo, Ding, Liu, Qiu, Liu \&
  Yu}]{zhang2022investigation}
\bibinfo{author}{Zhang, Y.}, \bibinfo{author}{Guo, B.}, \bibinfo{author}{Ding,
  Y.}, \bibinfo{author}{Liu, J.}, \bibinfo{author}{Qiu, C.},
  \bibinfo{author}{Liu, S.}, \& \bibinfo{author}{Yu, Z.}
  (\bibinfo{year}{2022}).
\newblock \bibinfo{title}{Investigation of the determinants for misinformation
  correction effectiveness on social media during covid-19 pandemic}.
\newblock {\it \bibinfo{journal}{Information Processing \& Management}\/},
  {\it \bibinfo{volume}{59}\/}, \bibinfo{pages}{102935}.
\bibitem[{Zhao et~al.(2020)Zhao, Yang, Lin \& Philip}]{zhao2020deep}
\bibinfo{author}{Zhao, Y.}, \bibinfo{author}{Yang, N.}, \bibinfo{author}{Lin,
  T.}, \& \bibinfo{author}{Philip, S.~Y.} (\bibinfo{year}{2020}).
\newblock \bibinfo{title}{Deep collaborative embedding for information cascade
  prediction}.
\newblock {\it \bibinfo{journal}{Knowledge-Based Systems}\/},  {\it
  \bibinfo{volume}{193}\/}, \bibinfo{pages}{105502}.
\bibitem[{Zhu et~al.(2021)Zhu, Dai, Chen \& Yin}]{zhu2021sampling}
\bibinfo{author}{Zhu, D.}, \bibinfo{author}{Dai, X.-Y.}, \bibinfo{author}{Chen,
  J.}, \& \bibinfo{author}{Yin, J.} (\bibinfo{year}{2021}).
\newblock \bibinfo{title}{Sampling informative context nodes for network
  embedding}.
\newblock {\it \bibinfo{journal}{Science China Information Sciences}\/},  {\it
  \bibinfo{volume}{64}\/}, \bibinfo{pages}{1--11}.

\end{thebibliography}


\end{document}


\title{Towards a performance characteristic curve for model evaluation: an application in information diffusion prediction: Supplementary materials}


\maketitle


\section{Supplementary experimental results}

\begin{figure}[!htb]
\begin{center}
\resizebox{1\textwidth}{!}{\includegraphics{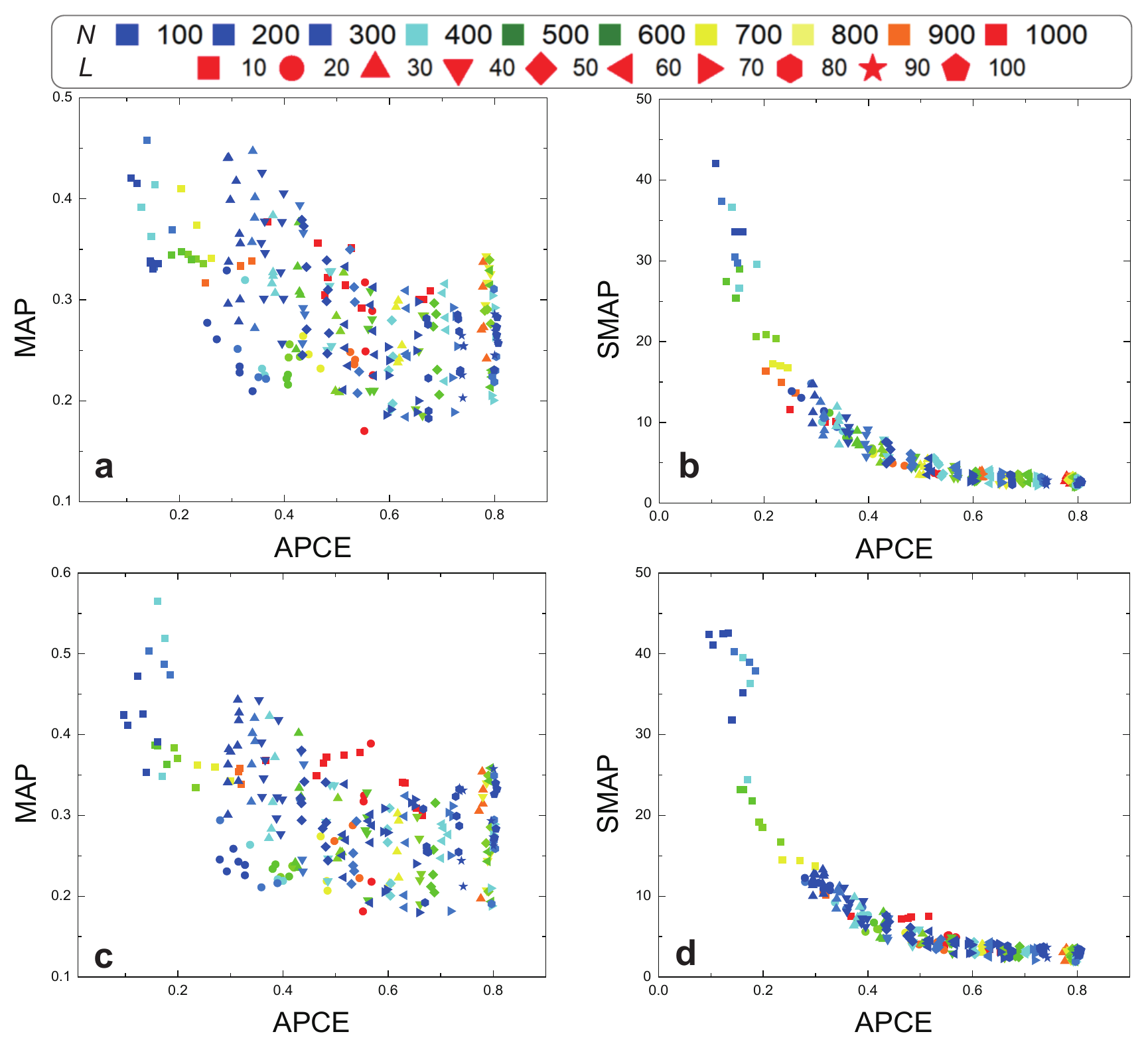}} 
\caption{The correlation patterns of APCE and the prediction performance of CDK prediction model. 
\textbf{a, b} The APCE-MAP pattern and APCE-SMAP pattern on sequence sets generated by LT model.
\textbf{c, d} The APCE-MAP pattern and APCE-SMAP pattern on sequence sets generated by SI model.}
\label{fig:LT-SI}
\end{center}
\end{figure}

Figure \ref{fig:LT-SI} shows the APCE-MAP and APCE-SMAP patterns of CDK model on the sequence sets generated by linear threshold model (LT) and susceptible-infectious model (SI). Compared to those on data sets generated by independent cascade model (IC) (Figure \textbf{3ab} in the main text), the experimental results are highly similar, indicating the scaling pattern proposed in this work is almost unaffected by the internal driven mechanism of the information spreading.

\begin{figure}[!htb]
\begin{center}
\resizebox{1\textwidth}{!}{\includegraphics{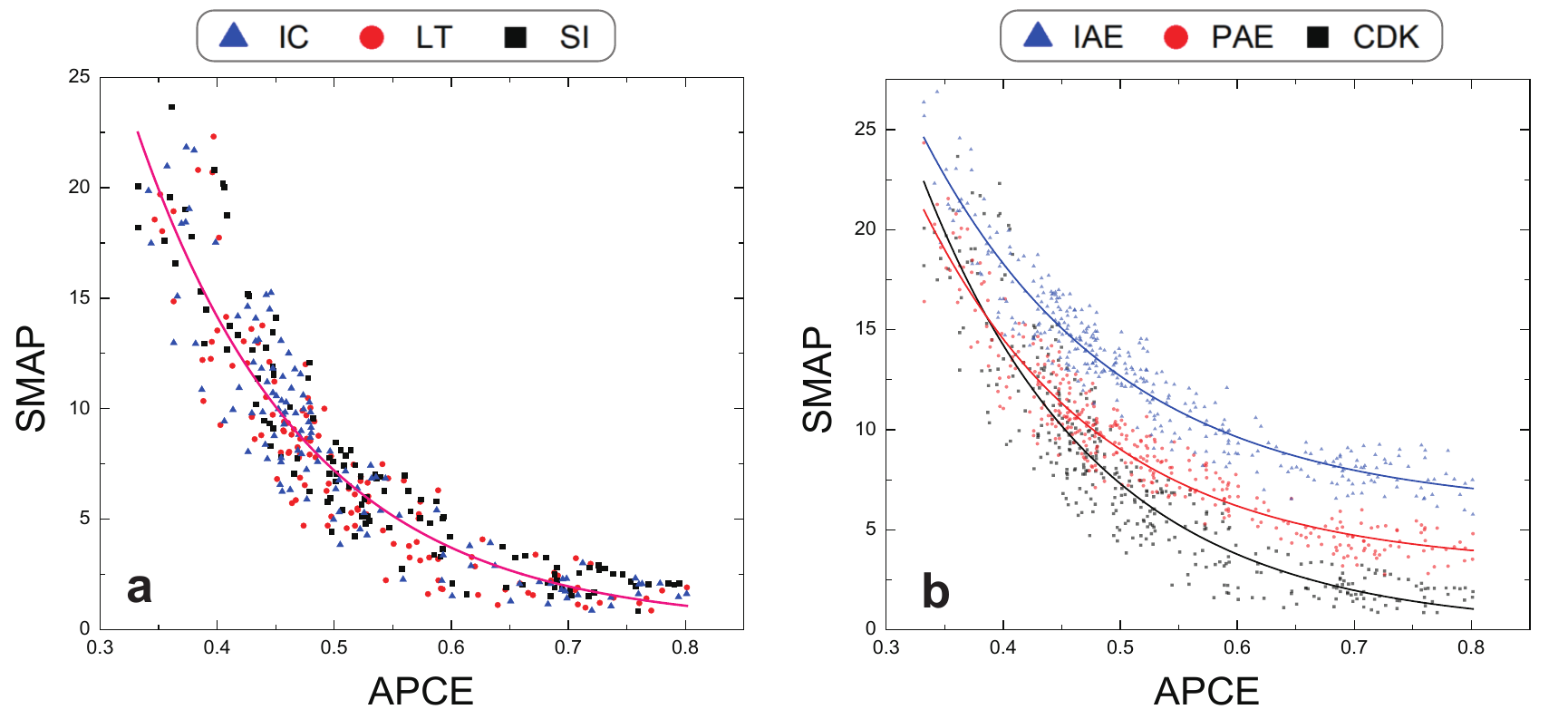}}
\caption{The APCE-SMAP scaling patterns on sequence sets generated on BA scale-free networks. 
\textbf{a} The APCE-SMAP scaling pattern of CDK on sequence sets generated by three different propagation mechanisms. The scaling curve is also fitted well by the decay exponential function as that on ER data set, but it does not coincide with the scaling curve on the ER data set, which might show the impact of network topology on the scaling pattern.
\textbf{b} The overall APCE-SMAP scaling patterns of CDK, PAE and IAE models. The patterns illustrate IAE outperforms PAE and PAE outperforms CDK obviously when APCE is high. But the weakness of CDK is getting smaller as the APCE decreases, and CDK even performs better than the PAE when APCE is low enough. This phenomenon is much similar to that on the ER data set, except that the CDK curve does not exceed the IAE curve when APCE is very low. We speculate that this is because the APCE values of the BA data set we generated do not reach a low enough value for CDK to show its merit (the lowest APCE is higher than 0.3 here, while the lowest APCE on ER data set is about 0.1).}
\label{fig:BA}
\end{center}
\end{figure}

Figure \ref{fig:BA} shows the scaling patterns on BA scale-free network data set. The performance characteristic curves on these patterns exhibit a similar trend to the curves on the patterns on ER random network data set. However, different from changing the spreading mechanism, changing network topology could lead to a change in the model's performance characteristic curve, which is still an exponential decay curve but with different parameters. It demonstrates the network topology could have a relatively greater impact on our model evaluation framework than the spreading mechanism.

\begin{figure}[!htb]
\begin{center}
\resizebox{1\textwidth}{!}{\includegraphics{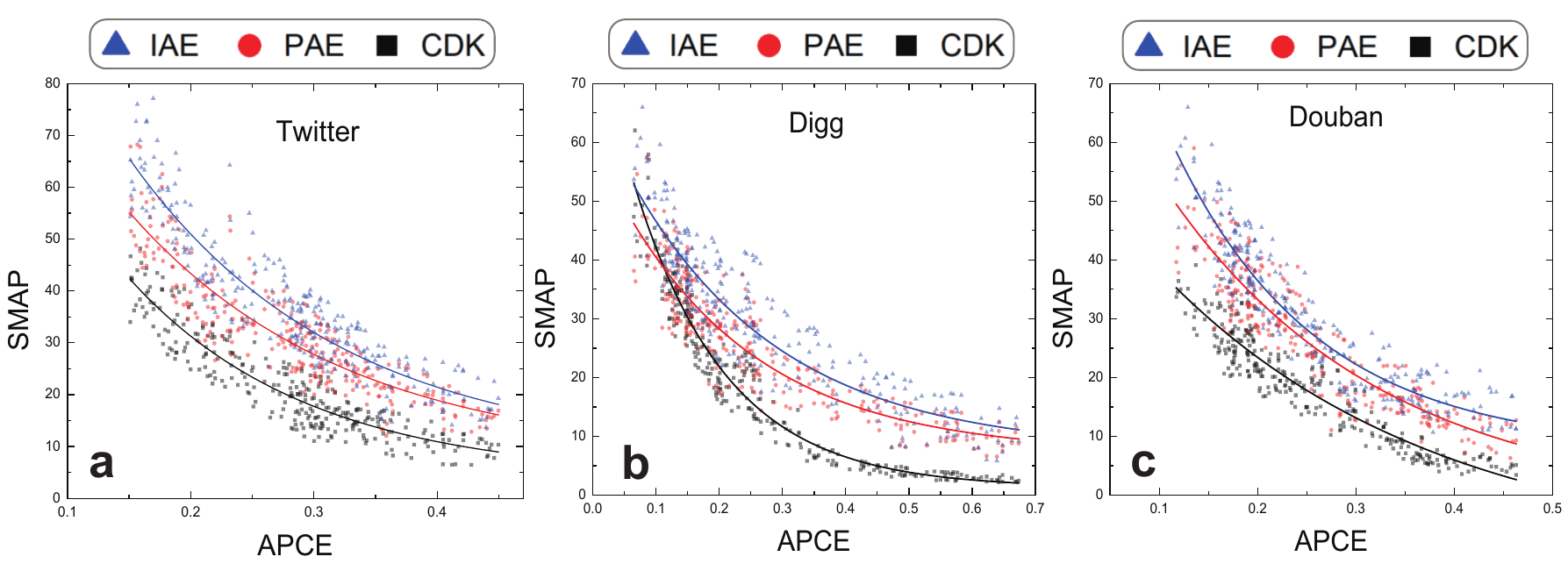}}
\caption{The APCE-SMAP scaling patterns of CDK, PAE and IAE models on empirical data. The patterns show more randomness in the distribution of nodes, but the scaling curves still illustrate IAE outperforms PAE and PAE outperforms CDK in most instances. Note that CDK performs better than PAE and close to IAE when APCE is less than 0.1 in \textbf{b}, illustrating CDK's merit in the easy prediction tasks as shown in the ER pattern (Figure \textbf{4} in main text) and BA pattern (Figure \ref{fig:BA}\textbf{b}) as well. This trend does not emerge in the Twitter pattern (\textbf{a}) and Douban pattern (\textbf{c}), which we speculate is caused by the limitations of the predictability of the data set. Due to the length of the diffusion sequence in Twitter and Douban data just ranging from 10 to 40 as we mentioned, the APCE values of these data sets are distributed in $(0.1,0.5)$, not so widely as those in Digg data. So the patterns on Twitter and Douban only show similar properties as the middle of patterns on Digg.}
\label{fig:emp_data_3models}
\end{center}
\end{figure}

Figure \ref{fig:emp_data_3models} shows the model evaluation results of CDK, PAE and IAE models on empirical data using the performance characteristic curves. In comparison to the results obtained on synthetic data, the results on empirical data exhibit greater randomness. Furthermore, due to the inherent characteristics of the three social media platforms in the empirical data, the prediction accuracy of the three models also varies across different data sets. However, overall, the comparative results of the models are consistent. In the majority of cases, the IAE model, which utilizes more embedding latent spaces, demonstrates the best prediction performance, aligning with the theoretical design principles of network-embedding-based prediction models. This validates the feasibility of the model evaluation approach based on the performance characteristic curve.

\begin{figure}[!htb]
\begin{center}
\resizebox{1\textwidth}{!}{\includegraphics{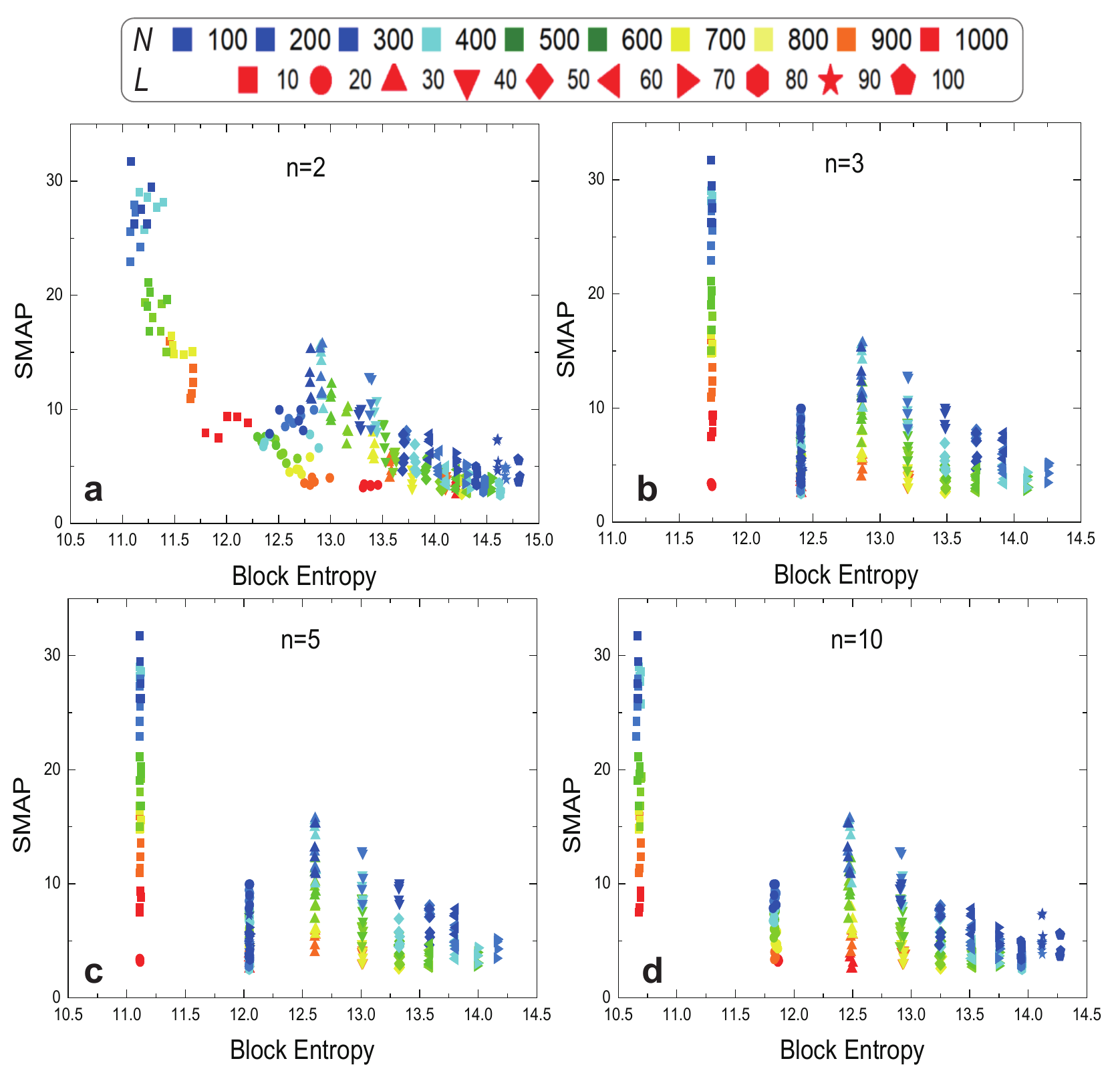}}
\caption{The scaling patterns between block entropy and SMAP on synthetic data. When $n=2$ in block entropy (Figure \textbf{a}), the block entropy can be seen as a simplified APCE that only considers continuous user pairs. Therefore, the obtained entropy value is relatively continuous, and the pattern shows an overall downward trend, similar to the APCE-SMAP pattern, but does not converge to an exponential function curve. When $n$ increases, the block entropy values are confined to only a few discrete points. Consequently, within these patterns, although the SMAP values demonstrate a decreasing trend with the increment of block entropy values, a scaling pattern that can be accurately fitted by a functional curve remains elusive (Figure \textbf{bcd}).}
\label{fig:block}
\end{center}
\end{figure}

Figure \ref{fig:block} shows the scaling pattern between block entropy and SMAP. To make a better comparison with APCE, the experiments are conducted on synthetic data and the SMAP is obtained by CDK model, same settings as the experiments in Section 4.2 in main text. The parameter $n$ in block entropy (see Equation 3 in main text) is adjusted to facilitate observation of the performance of block entropy under different $n$. When $n=2$, block entropy examines the frequency of every two consecutive element combinations. In this case, block entropy can be considered as a simplified version of APCE, with a scaling pattern that is relatively similar to the APCE-SMAP pattern. When the value of $n$ increases, the probability of multiple occurrences of blocks of $n$ consecutive elements in the sequence decreases. In fact, in many cases, these n-blocks only appear once in the sequence set. This resulting in only a few discrete values for block entropy, making it even more difficult to obtain clear and comprehensive scaling patterns.

\clearpage


